\documentclass[10pt,twocolumn,letterpaper]{article}

\usepackage[pagenumbers]{iccv} 
\usepackage{adjustbox}
\usepackage{caption}
\usepackage{color}
\usepackage{xcolor}
\usepackage[dvipsnames]{xcolor}

\usepackage{multirow}

\definecolor{iccvblue}{rgb}{0.21,0.49,0.74}
\usepackage[pagebackref,breaklinks,colorlinks,allcolors=iccvblue]{hyperref}

\def\paperID{2613} 
\def\confName{ICCV}
\def\confYear{2025}

\title{ModalTune: Fine-Tuning Slide-Level Foundation Models with Multi-Modal Information for Multi-task Learning in Digital Pathology}

\author{
  Vishwesh Ramanathan$^{1,2}$\thanks{Equal contributions}\quad
  Tony Xu$^{1,2}$\footnotemark[1]\quad
  Pushpak Pati$^{4}$\quad
  Faruk Ahmed$^{3}$\quad
  Maged Goubran$^{1,2}$\\
  Anne L. Martel$^{1,2}$\\
  $^1$Sunnybrook Research Institute, Canada\\
  $^2$University of Toronto, Canada\\
  $^3$Google Research, USA\\
  $^4$Independent Researcher\\
}

\begin{document}
\maketitle
\begin{abstract}
Prediction tasks in digital pathology are challenging due to the massive size of whole-slide images (WSIs) and the weak nature of training signals. Advances in computing, data availability, and self-supervised learning (SSL) have paved the way for slide-level foundation models (SLFMs) that can improve prediction tasks in low-data regimes. However, current methods under-utilize shared information between tasks and modalities. To overcome this challenge, we propose ModalTune, a novel fine-tuning framework which introduces the Modal Adapter to integrate new modalities without modifying SLFM weights. Additionally, we use large-language models (LLMs) to encode labels as text, capturing semantic relationships across multiple tasks and cancer types in a single training recipe. ModalTune achieves state-of-the-art (SOTA) results against both uni-modal and multi-modal models across four cancer types, jointly improving survival and cancer subtype prediction while remaining competitive in pan-cancer settings. Additionally, we show ModalTune is generalizable to two out-of-distribution (OOD) datasets. To our knowledge, this is the first unified fine-tuning framework for multi-modal, multi-task, and pan-cancer modeling in digital pathology. Our code is publicly available at \url{ https://github.com/martellab-sri/ModalTune}.

\end{abstract}

\section{Introduction}
\label{sec:intro}

\begin{figure}[t]
  \centering
  
   \includegraphics[width=1.0\linewidth]{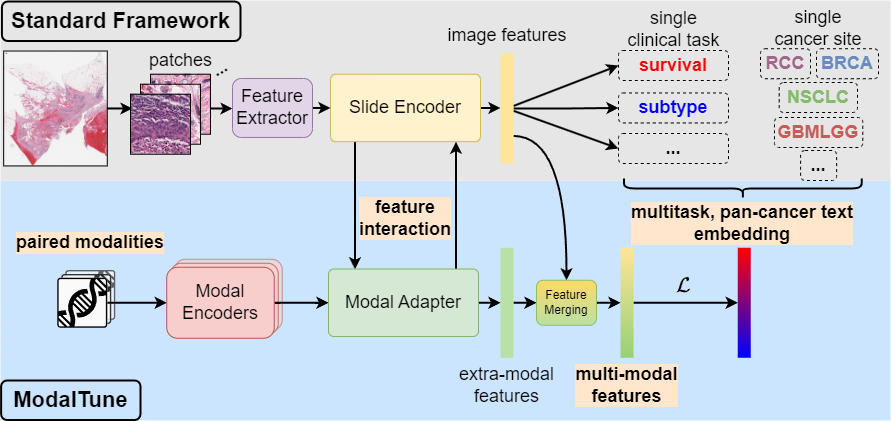}

   \caption{Unlike standard fine-tuning approaches that focus on a single task, modality, and cancer site, our novel fine-tuning framework, ModalTune, leverages shared information across multiple tasks, modalities, and cancer sites to interface with SLFMs.}
   \label{fig:brief_summary}
\end{figure}

Prediction tasks in histopathology are challenging due to the massive size and complexity of WSIs, which can contain critical diagnostic information for tumor staging, subtyping, and prognosis assessment. These predictive signals are often at the slide-level, requiring weakly-supervised learning approaches. Multiple Instance Learning (MIL) has become a key method for WSI analysis, dividing WSIs into patches, extracting features using pre-trained models, and aggregating them into slide-level representations through various pooling methods \cite{ilse2018attention,shao2021transmil,patchgcn,zheng2022graph,thandiackal2022differentiable,ramanathan2024ensemble}. With the development of large patch-level foundation models (FMs) \cite{ciga2022self, wang2022transformer, Chen2024TowardsPathology, lu2024visual}, feature extraction has improved, thereby enhancing the performance of MIL in slide-level tasks \cite{Campanella2024}.

Recent advances in computational resources, data availability, and algorithm efficiency have led to the development of SLFMs, which involve training MIL aggregators (`slide encoders') using SSL techniques. These SSL techniques include masked image modeling \cite{He2022MaskedLearners, xu2024whole}, contrastive learning with transcriptomics \cite{jaume2024transcriptomics, Vaidya2025Molecular-drivenPathology}, pathology reports \cite{Ding2024MultimodalPathology, Ahmed2024PathAlign:Histopathology}, and multiple stains \cite{jaume2024multistain}, marking a shift from patch-level FMs to robust SLFMs. Yet, many domain-specific variations may still not be captured during pre-training, such as differences in tissue preparation across clinical centers, OOD morphological characteristics, and the presence of complex or rare diseases. This makes fine-tuning of SLFMs a more effective strategy for deployment on new datasets.

Current frameworks for deploying SLFMs are limited to fine-tuning them on a single task and cancer site while taking only WSIs as input (\cref{fig:brief_summary}). They fail to fully leverage rich datasets, which may include other modalities like clinical records, reports, and genomics, as well as extensive slide annotations like cancer subtypes, tumor staging, prognosis, and biomarker presence. Existing frameworks face two key limitations when applied in this setting: (1) Existing SLFMs use only WSIs as input during inference, overlooking other modalities. While these models may have incorporated other modalities for training, there are significant benefits to leveraging them during inference to capitalize on multi-modal insights \cite{jaume2024modeling,Chen2021MultimodalImages}. (2) Current frameworks often emphasize single-task learning, overlooking interconnected multi-task information that could enhance SLFMs' fine-tuning performance and generalization \cite{crawshaw2020multi}. These challenges result in suboptimally fine-tuned SLFMs.  

To address these limitations, we propose two enhancements that transform SLFMs into multi-modal and multi-task learners while preserving their foundational knowledge: (1) We introduce the \textit{Modal Adapter}. Inspired by recent work in NLP and natural images \cite{pfeiffer2020adapterhub, Chen2022VisionPredictions}, this module interfaces with slide-encoders in histopathology to integrate information from additional modalities without altering the primary model’s weights. (2) We introduce \textit{text-based multi-task learning}. We convert downstream tasks (\eg subtype classification, risk prediction) into text and use LLMs to represent them as text vectors in a unified embedding space. We tune our models to predict these text embeddings using a single loss function, regardless of the task type or number, enabling the model to leverage semantic relationships between entities within and across tasks.

Our overall method called \textit{ModalTune} is a fine-tuning framework designed to fully leverage the potential of SLFMs and pathology datasets. Our main contributions are:
\begin{itemize}
    \item \textbf{Modal Adapters}: We propose Modal Adapters to continually inject multi-modal information into slide encoders without altering their weights. It can interface with any transformer-based slide encoder and dynamically scale to new modalities.
    \item \textbf{Multi-Task Learning}: We simplify multi-task learning by mapping tasks into a shared embedding space using LLMs, leveraging complementary semantic and inter-task relationships. Our formulation optimizes a single loss function regardless of the number or type of tasks.
    \item \textbf{Pan-Cancer Fine-Tuning}: Our framework integrates multiple cancer sites, resulting in a single resource-efficient, pan-cancer model that generalizes across tasks and sites. We show this approach enables smaller datasets to leverage key insights from larger ones. 
\end{itemize}
To the best of our knowledge, we propose the first fine-tuning framework for unified approach for multi-modal, multi-task, and pan-cancer modeling in digital pathology.
 
We validate our approach on four TCGA datasets \cite{tomczak2015review}: Breast Invasive Carcinoma (BRCA), Non-small Cell Lung Carcinoma (NSCLC), Glioblastoma Multiforme and Low-Grade Glioma (GBMLGG), and Renal Cell Carcinoma (RCC), covering 4,099 slides and 3,418 patients. Using Prov-Gigapath \cite{xu2024whole} (referred to as Gigapath) as our SLFM, our model outperforms uni-modal and multi-modal baselines in cancer subtype and survival prediction, achieving higher balanced accuracy and C-indices than the fine-tuned uni-modal Gigapath (+4.6\%, +9.6\%), its multi-modal counterpart (+2.8\%, +7.0\%), and the best baseline (+1.4\%, +1.9\%). We further tested our model on two smaller OOD datasets: Colon and Rectum Adenocarcinoma (COADREAD) and Bladder Urothelial Carcinoma (BLCA), where it retained generalizability, performed competitively with the supervised baseline and outperformed its fine-tuned counterpart by a large margin (+25\%, +8.3\%).
\section{Related Work}
\label{sec:rel work}
This section reviews related works on WSI-text alignment, multi-modal and multi-task methods in digital pathology.

\noindent \textbf{WSI-Text Alignment}: Recent vision-language models (e.g. CLIP \cite{Radford2021LearningSupervision}, CoCa \cite{yu2022coca}, Llava \cite{liu2024visual}, and BLIP-2 \cite{li2023blip}) have advanced text alignment at the patch-level in digital pathology \cite{lu2024visual, ZhouPathM3:Captioning, lu2024multimodal}. More recently, studies have aligned WSIs with slide-level diagnostic reports, with PRISM \cite{Shaikovski2024PRISM:Histopathology} using attention mechanisms and CoCa \cite{yu2022coca}, and Gigapath \cite{xu2024whole} leveraging masked autoencoder pre-training \cite{He2022MaskedLearners} and contrastive learning on pathology reports. TITAN \cite{Ding2024MultimodalPathology} enhances WSI-text alignment by first aligning slide embeddings with synthetic region-level captions, then refining them using slide-level pathology reports for improved representation. PathAlign \cite{Ahmed2024PathAlign:Histopathology} under the BLIP-2 \cite{li2023blip} framework integrates slide encoder with frozen LLMs for report generation, WSI classification, and text retrieval.

While our work shares similarities with contrastive learning-based WSI-text alignment, it introduces key differences. Existing methods rely on large-scale datasets, pathology reports, and large batch sizes, limiting their adaptability. In contrast, our approach aligns more closely with typical MIL frameworks, enabling fine-tuning in a single-batch setting using widely available synoptic data, hence making it adaptable to any dataset.

\noindent \textbf{Multi-Modal Models}: Multiple studies in digital pathology show that integrating WSIs with clinical data—such as RNA transcriptomics \cite{jaume2024modeling, chen2020pathomic, jin2023gene, Chen2021MultimodalImages}, copy number alterations (CNA) \cite{Ding2023Pathology-and-genomicsPrediction, Chen2021MultimodalImages}, MRI scans \cite{pei2021hybrid}, or clinical captions \cite{ZhouPathM3:Captioning}—captures complementary biological and contextual insights, improving prediction over single-modality models. Multi-modal fusion in digital pathology employ various strategies to integrate heterogeneous data. Early fusion combines raw input features before processing, with initial methods using concatenation or pooling \cite{huang2020fusion}. Inspired by advances in Visual Question Answering field, recent approaches leverage cross-attention mechanisms \cite{Chen2021MultimodalImages,jaume2024modeling, ZhouPathM3:Captioning}. Late fusion aggregates predictions from modality-specific models using methods like concatenation \cite{mobadersany2018predicting, 7950668}, Kronecker products \cite{chen2020pathomic}, or transformer-based fusion \cite{Xu_2023_ICCV}.

Our Modal Adapters, inspired by ViT Adapters \cite{Chen2022VisionPredictions}, differ by performing cross-attention between modalities at multiple hidden layers, followed by summation-based fusion to generate the final feature representations.

\noindent \textbf{Multi-Task Learning}: Multi-task learning (MTL) optimizes related tasks from the same input by leveraging shared representations, enhancing data efficiency, generalizability, and reducing overfitting \cite{crawshaw2020multi}, making it especially useful in digital pathology which has co-related tasks like cancer subtyping, survival, and biomarker prediction. Early work at the patch level by Li et al. \cite{li2020multi} used histopathology patches for breast cancer classification and grading, while Simon et al. \cite{graham2023one} proposed a unified model for cell and tissue segmentation and classification. At the WSI-level, Weng et al. \cite{weng2019multimodal} introduced MTL for predicting multiple slide-level metadata, Gao et al. \cite{gao2023semi} developed a framework for cancer region detection and subtyping, and Zhao et al. \cite{Zhao2023MulGT:Analysis} proposed a multi-task graph transformer for tumor typing and staging. More recently, Shao et al. \cite{Shao2024Multi-InstanceImages} explored MTL for cancer subtpying, survival prediction, and transcriptomic profiling of gene TP53.

Unlike prior works that optimize multiple task-specific loss functions, our approach uses LLMs to unify tasks into a single representation, requiring only a single loss function for training. This offers greater simplicity and flexibility, making it easier to adapt to diverse or additional task sets.

\section{Method}

\begin{figure*}[t]
  \centering
   \includegraphics[width=0.87\linewidth]{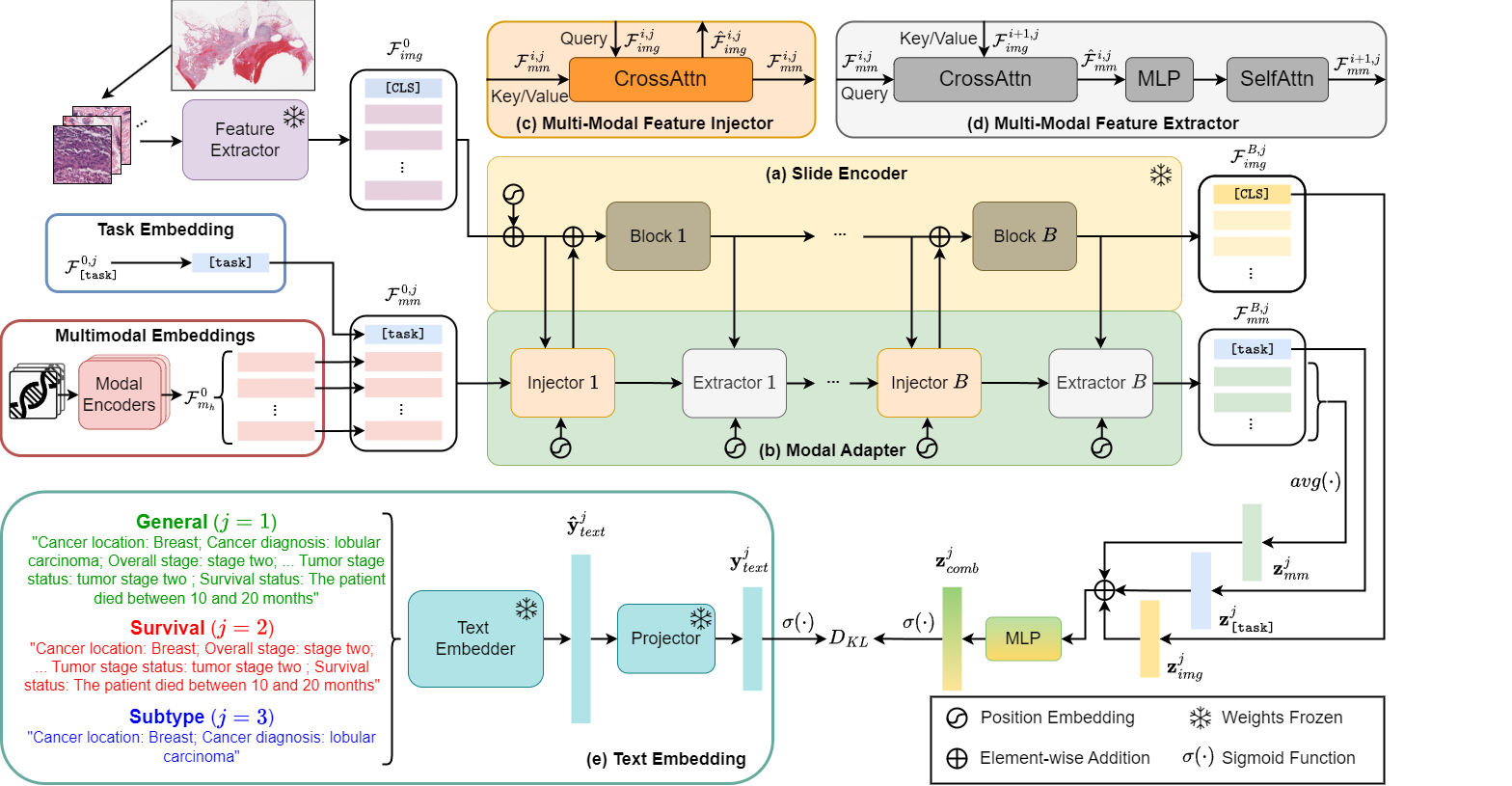}
   \caption{Overview of ModalTune. Built on a frozen pre-trained slide encoder (\textbf{a}), our Modal Adapter (\textbf{b}) integrates extra-modal features via the Multi-Modal Feature Injector (\textbf{c}) and Extractor (\textbf{d}). A text embedding module (\textbf{e}) unifies multiple tasks and cancer sites, while the enriched image, task, and modal embeddings are fused for training, ensuring a robust multi-modal, multi-task representation.}
   \label{fig:main_methods}
\end{figure*}

We introduce ModalTune, illustrated in \cref{fig:main_methods}, starting with conventional pretrained slide encoder (\cref{sec:slide_enc}), and feature extraction from additional paired modalities (\cref{sec:modal_enc}). We then present the Modal Adapter (\cref{sec:modal_adapt}) that injects additional modality features into the slide encoder. Finally, we describe the text embedding process (\cref{sec:text_embed}) for unifying tasks and cancer sites into a single embedding space and explain the overall training objective (\cref{sec:objective}).


\subsection{Slide Encoder} \label{sec:slide_enc}

We build on the SLFM Gigapath \cite{xu2024whole}, chosen for its strong performance, computational efficiency, and publicly available pre-trained weights. Gigapath effectively models slide embeddings by leveraging all patches and scales well in scenarios involving multiple slides with a large number of patches, enabling analysis without excessive overhead. This balance of efficiency and coverage makes it a practical choice for our study, even though more recent SLFMs may achieve stronger benchmark performance. Importantly, our framework remains fully adaptable to any transformer-based SLFM. For experiments on ModalTune applied to recent SLFM like TITAN \cite{Ding2024MultimodalPathology}, please refer to Supp. \cref{sec:supp_titan}.

Slide encoders extract WSI features using a two-step process. First, a WSI $\mathbf{X}_{img}$ is divided into $N_{img}$ patches, and Gigapath's pre-trained patch encoder extracts $D_{img}$ dimensional features, resulting in $\mathbf{\hat{X}}_{img} \in \mathbb{R}^{N_{img} \times D_{img}}$. The pre-trained Gigapath model, an $L$-layer LongNet \cite{Wang2023WhenPathology}, facilitates interactions among patch embeddings. Before further processing, these patch features pass through a linear layer to produce $\mathcal{F}_{img}^0 \in \mathbb{R}^{N_{img} \times D}$, where $D$ is the LongNet embedding size. For ease of notation in \cref{sec:modal_adapt}, the $L$ LongNet layers are divided into $B$ blocks, each with $L/B$ layers. Following the Gigapath model, we use the frozen pre-trained \texttt{[CLS]} token from the slide encoder to aggregate patch embeddings into a single representation. Thus, after passing through the $i$-th LongNet block, the patch embeddings are $\mathcal{F}_{img}^i \in \mathbb{R}^{(N_{img}+1) \times D}$, for $i \in \{1, 2, ..., B\}$.

\subsection{Modal Encoders} \label{sec:modal_enc}
Modal encoders are employed to extract features from each additional modality, which will be used to inform the pre-trained slide encoder. We define $M$ generic new modalities, denoted as $m_h$ for $h \in \{1, 2, \ldots, M\}$, with input data $\mathbf{X}_{m_h}$ paired with the WSI $\mathbf{X}_{img}$. Each modality's input is independently encoded by the modal encoders, producing a feature set $\mathcal{F}_{m_h}^0 \in \mathbb{R}^{N_{m_h} \times D}$, where $N_{m_h}$ is the token length for each modality. The final multi-modal feature set is formed by concatenating features from all modalities, resulting in $\mathcal{F}_{mm}^0 \in \mathbb{R}^{(N_{m_1} + N_{m_2} + \ldots + N_{m_M}) \times D}$.

\subsubsection{Transcriptomics Encoder}

Bulk transcriptomics data is increasingly available and provides complementary insights on cancer biology to WSIs \cite{chen2020pathomic, Chen2021MultimodalImages, jaume2024modeling}, allowing it to effectively predict patient survival \cite{elmarakeby2021biologically, baul2024integrating}. Thus, we consider it our primary additional modality in this work. Other modalities, such as clinical data (see Supp. \cref{sec:supp_multimodal}), additional ``omics,'' imaging techniques, or domain-specific priors, can in principle be integrated by concatenating their extracted feature tokens. However, further exploration of these extensions is left for future work.

The transcriptomics encoder compresses bulk transcriptomics data, $\mathbf{X}_{t} \in \mathbb{R}^{N_g}$, containing $N_g$ genes into a set of tokens. We begin by dividing the genes into $N_{gp}$ gene pathways, which are encoded using a sparse multi-layer perceptron (S-MLP) \cite{elmarakeby2021biologically} into pathway features, $\tilde{\mathcal{F}}_{gp} \in \mathbb{R}^{N_{gp} \times D_{gp}}$, as described in \cite{jaume2024modeling}. To merge intra- and inter-pathway features, we use the MLP-mixer \cite{Tolstikhin2021MLP-Mixer:Vision}, encoding pathway features to produce $\mathcal{F}_{gp} \in \mathbb{R}^{N_{gp} \times D}$. Finally, a linear projection reduces the number of pathways to $N_t$, enhancing downstream tractability. The encoder finally outputs a compressed set of pathway tokens, $\mathcal{F}_t^0 \in \mathbb{R}^{N_{t} \times D}$.

\subsection{Modal Adapter} \label{sec:modal_adapt}

To integrate multi-modal features while taking advantage of pre-trained slide encoders, we introduce the Modal Adapter. Our adapter introduces informative cross-modality features into slide encoders, drawing inspiration from the ViT-Adapter \cite{Chen2022VisionPredictions}. Specifically, it merges compressed pathway features from the transcriptomics encoder with image features from the slide encoder, enriching the network with meaningful multi-modal representations.


\subsubsection{Task Prompt} \label{sec:task_prompt}

The Modal Adapter requires distinct feature sets tailored to each downstream task. To facilitate feature interactions and attention operations, we introduce a learnable \texttt{[task]} prompt. This consists of a set of learnable vectors, $\mathcal{F}_{\texttt{[task]}}^{0, j} \in \mathbb{R}^{1 \times D}$, for each downstream task $j \in \{1, 2, \ldots, T\}$, capturing task-specific biases across $T$ tasks. This prompt is concatenated with $N_t$ modal tokens derived from compressed pathways in our approach, to create the complete set of multi-modal features: $\mathcal{F}_{mm}^{0, j} \in \mathbb{R}^{(N_{t}+1) \times D}$.

\subsubsection{Multi-modal Feature Interaction}

Feature interaction in the Modal Adapter is governed by \textit{Multi-modal Feature Injector} modules, which inject multi-modal features into image features, and \textit{Multi-modal Feature Extractor} modules, which extract image features to inform the features from extra modalities. 

\noindent \textbf{Multi-Modal Feature Injectors} 
are formulated as standard cross-attention, $\text{CrossAttn}(\cdot)$ \cite{vaswani2017attention}, with query being the WSI features, 
$\mathcal{F}_{img}^{i, j} \in \mathbb{R}^{(N_{img}+1) \times D}$, and key and value being the multi-modal features, $\mathcal{F}_{mm}^{i, j} \in \mathbb{R}^{(N_{t}+1)\times D}$. The Injector for the $i$-th LongNet block and $j$-th task is given as,
\begin{equation}
  \hat{\mathcal{F}}_{img}^{i, j} = \mathcal{F}_{img}^{i, j} + \gamma^i \text{CrossAttn}(\text{LN}(\mathcal{F}_{img}^{i, j}), \text{LN}(\mathcal{F}_{mm}^{i, j}))
  \label{eq:injector}
\end{equation}
where, $\text{LN}(\cdot)$ is LayerNorm \cite{ba2016layernormalization}, and $\gamma^i \in \mathbb{R}^D$ is a learnable vector controlling how the original image features are adjusted by the cross-attention layer. Based on \cite{Chen2022VisionPredictions}, we initialize $\gamma^i$ with $\mathbf{0}$ to allow multi-modal features to be \textit{gradually} introduced into the pre-trained slide encoder during training. The image features from the next LongNet block, $\mathcal{F}_{img}^{i+1, j}$, are obtained by passing the post-injection image features, $\hat{\mathcal{F}}_{img}^{i, j}$ through the $i$-th LongNet block. 

\noindent \textbf{Multi-modal Feature Extractors} introduce image features \textit{back into} multi-modal features. It is implemented using cross-attention, followed by an MLP and a self-attention operation $\text{SelfAttn}(\cdot)$ to model interactions between different modalities and task tokens. Here, the query is $\mathcal{F}_{mm}^{i, j}$, and the key and value are the image features after passing through a LongNet block, $\mathcal{F}_{img}^{i+1, j}$, as shown in \cref{eq:extractor1} and \cref{eq:extractor2}.
\begin{equation}
  \hat{\mathcal{F}}_{mm}^{i, j} = \mathcal{F}_{mm}^{i, j} + \text{CrossAttn}(\text{LN}(\mathcal{F}_{mm}^{i, j}), \text{LN}(\mathcal{F}_{img}^{i+1, j}))
  \label{eq:extractor1}
\end{equation}
\begin{equation}
  \mathcal{F}_{mm}^{i+1, j} = \text{SelfAttn}\left(\hat{\mathcal{F}}_{mm}^{i, j} + \text{MLP}(\text{LN}(\hat{\mathcal{F}}_{mm}^{i, j}))\right)
  \label{eq:extractor2}
\end{equation}

\subsubsection{Output Feature Representation}
\label{sec:output}
We extract the representation from the final LongNet block, $\mathcal{F}_{img}^{B, j} \in \mathbb{R}^{(N_{img}+1) \times D}$, using the \texttt{[CLS]} token to get the \textit{image} feature, $\mathbf{z}_{img}^j \in \mathbb{R}^{D}$. We use the final multi-modal features $\mathcal{F}_{mm}^{B, j} \in \mathbb{R}^{(N_{t}+1) \times D}$, indexed by the \texttt{[task]} token, to obtain the task-specific vector, $\mathbf{z}_{\texttt{[task]}}^j \in \mathbb{R}^{D}$. We also average the multi-modal features across the $N_t$ compressed pathway tokens to get a modality-specific feature, $\mathbf{z}_{mm}^j$.
Here, $\mathbf{z}_{\texttt{[task]}}^j$ encapsulates task-specific information enriched with image and extra-modality features, while $\mathbf{z}_{mm}^j$ captures modality-specific information contextualized by image and task features. We average modality features to form $\mathbf{z}_{mm}^j$ because the $\mathbf{z}_{\texttt{[task]}}^j$ primarily captures enriched task information, making it inadequate to represent modality features independently. Therefore, excluding the modality representation $\mathbf{z}_{mm}^j$ would lead to an incomplete feature set.
The final output representation, $\mathbf{z}_{comb}^j \in \mathbb{R}^{D_{final}}$ combining all features tailored for the $j$-th task is produced as, 
\begin{equation}
  \mathbf{z}_{comb}^j = \text{MLP}(\text{LN}(\mathbf{z}_{img}^j + \mathbf{z}_{mm}^j + \mathbf{z}_{\texttt{[task]}}^j))
  \label{eq:final_out}
\end{equation}
This vector can be used alongside classical machine learning methods on downstream tasks. Importantly, the Modal Adapter does not update the slide encoder weights, taking better advantage of pre-trained SLFMs and reducing the computational cost of our methodology. 

\subsection{Text Embedding} \label{sec:text_embed}

We encode tasks using text to unify tasks, datasets, and cancer sites into a single embedding space, eliminating the need for complex loss functions. We generate text embeddings by converting clinical tabular data into text (\cref{fig:main_methods}) and processing it with a pre-trained text encoder. Here, we use the text encoder from CONCH \cite{lu2024visual}, pre-trained on histology captions. Since different tasks require unique feature interactions (\cref{sec:task_prompt}), we create a `general' task ($j = 1$) text embedding with all clinically relevant information and `task-specific' ($j > 1$) embeddings tailored to each downstream task. The text for the $j$-th task is encoded into a vector, $\mathbf{\hat{y}}_{text}^j \in \mathbb{R}^{D_{text}}$, and passed through a fixed, randomly-initialized \textit{Projector} to produce the final text embedding, $\mathbf{y}_{text}^j \in \mathbb{R}^{D_{final}}$. The Supplementary provides further details on text construction (\cref{sec:supp_Text}) and analysis (\cref{sec:supp_textembeddinganalysis}).

\subsection{Training Objective} \label{sec:objective}
We aim to replicate the text embedding distribution to capture both task-specific information and semantic relationships. To achieve this, we use Kullback–Leibler (KL) divergence, following \cite{li2024promptkd}, who demonstrated its advantages over CLIP \cite{radford2021learning} and other distillation losses.
ModalTune is optimized to align the multi-modal embedding $\mathbf{z}_{comb}^j$ for task $j$ with its corresponding text embedding $\mathbf{y}_{text}^j$. The alignment is achieved by minimizing the KL divergence loss, averaged across all tasks, as shown in \cref{eq:loss}, where $\sigma(\cdot)$ denotes softmax function.

\begin{equation}
\small
  \mathcal{L} = \frac{1}{T} \sum_{j=1}^{T} D_{KL}\left[\sigma \left(\frac{\mathbf{z}_{comb}^j}{||\mathbf{z}_{comb}^j||_2}\right)\, \Big|\Big|\, \sigma \left(\frac{\mathbf{y}_{text}^j}{||\mathbf{y}_{text}^j||_2}\right)\right]
  \label{eq:loss}
\end{equation}

\section{Experiments and Results}

\subsection{Datasets}

We utilized four TCGA datasets \cite{tomczak2015review}—BRCA, NSCLC, GBMLGG, and RCC—to train and evaluate ModalTune for cancer subtype and survival prediction, covering 4,099 slides and 3,418 patients. COADREAD and BLCA datasets were used as OOD data to test model generalizability. Data was split at patient-level into [68\%, 12\%, 20\%] for training, validation, and testing, stratified by cancer subtypes (\cref{tab:dataset}). Transcriptomics data were sourced from the Xena database \cite{goldman2020visualizing} and preprocessed similar to SurvPath \cite{jaume2024modeling}, resulting in 331 pathways with 4,987 genes. For subtyping, subtypes were grouped into broader categories based on OncoTree code definitions \cite{kundra2021oncotree}, with rare subtypes ($<25$ cases) assigned to a `Rare-set' used only for training with text embeddings. Baseline models cannot use this class due to insufficient occurrences for effective label-based training. Exact groupings are detailed in the Supp. \cref{sec:supp_classgroupings}. For each patient, we extracted $256 \times 256$ patches from WSI foreground regions (via thresholding in HSV colorspace) at $0.5 \mu m/\textit{pixel}$. Since outputs are generated on a patient-wise basis, patches from multiple WSIs were concatenated into a single bag. Features of size $D_{img} = 1536$ were extracted from each patch using Gigapath's DINOv2 \cite{oquab2023dinov2} pre-trained feature extractor.

\begin{table}
\begin{adjustbox}{width=0.95\columnwidth} 
\small
\begin{tabular}{lcccc|cc}
\hline
&\textbf{BRCA} & \textbf{GBMLGG} & \textbf{NSCLC} & \textbf{RCC} & \textbf{COADREAD} & \textbf{BLCA}\\
\hline
Train cases & 747 & 401 & 667 & 574 & 273 & 256  \\
Val cases & 113 & 71 & 100 & 102 & 38 & 46\\
Test cases & 188 & 118 & 167 & 170 & 64 & 76\\
\hline
\# events & 146 & 200 & 362 & 219 & 85 & 171\\
\# censorships & 902 & 390 & 572 & 627 & 290 & 207\\
\hline
Rare-set & 111 & 0 & 103 & 0 & 58 & 0\\
Class 0 & 746 & 102 & 378 & 271 & 248 & 312\\
Class 1 & 191 & 488 & 453 & 510 & 69 & 66\\
Class 2 & - & - & - & 65 & - &  -\\
\hline
\end{tabular}
\end{adjustbox}
\caption{Dataset statistics. Note: `Rare-set' denotes cancer subtypes not considered for training due to their relative rarity.}  
\label {tab:dataset}
\end{table}

\subsection{Baseline Setup} 

Baselines are trained in a single-task, single-cancer manner for classification and survival tasks. `WSI' baselines use only WSIs, including popular models like ABMIL \cite{ilse2018attention} and TransMIL \cite{shao2021transmil}. We also assess the standalone SLFM, Gigapath \cite{xu2024whole}, with full end-to-end tuning. `Genomics' baselines rely on bulk transcriptomics, including MLP, SNN \cite{klambauer2017self}, and S-MLP \cite{elmarakeby2021biologically}. We also test ModalTune's transcriptomics encoder, integrating an MLP-mixer \cite{Tolstikhin2021MLP-Mixer:Vision} with S-MLP. `Multi-modal' baselines combine image and genomic data, including cross-attention-based early fusion methods like MCAT \cite{Chen2021MultimodalImages} and SurvPath \cite{jaume2024modeling}. We also evaluate late fusion methods, merging image-only baselines with MLP-mixers via concatenation \cite{mobadersany2018predicting} and Kronecker product \cite{chen2020pathomic}.

\subsection{ModalTune Setup} 

We configure ModalTune for $T=3$ tasks: general, survival, and subtype prediction. Gene pathways are linearly projected to $N_t = 64$ dimensions for tractability, and the final output feature dimension is $D_{final}=256$. Training runs for $30$ epochs with a $0.0001$ learning rate and batch size of $1$. Additional details are in Supp. \cref{sec:supp_Hyperparameters}.

After training, we evaluate its feature quality for classification and survival prediction. Using the General ($j=1$) task prompt, we extract features from both training and test datasets, selecting the best training epoch based on validation classification performance. For final alignment, we use linear probing (LP), fitting a linear classifier \cite{scikit-learn} for cancer subtypes and a Cox Proportional Hazards (CPH) model \cite{Davidson-Pilon2019} for survival prediction. All ModalTune results are \textit{multitask}, utilizing a single feature set for both tasks. `ModalTune' experiments are trained per cancer site and applied to downstream tasks within the same site, while `ModalTune Pan-Cancer' pools all cancer sites during training and fits separate classifiers or CPH models per site for prediction.

\subsection{Generalization Study Setup} \label{sec:generalization_study}
To assess the generalizability of our pipeline, we apply fine-tuned models to unseen TCGA datasets: COADREAD and BLCA. `ModalTune Pan-Cancer' applies the trained in-domain model to extract features on OOD datasets. `ModalTune' extracts features using the model trained on BRCA. `Gigapath Surv./Cls. (cat)' applies the respective single-task models directly tuned on BRCA and removes the final prediction layer to extract WSI features. We perform linear probing after feature extraction for all these experiments. The final `Gigapath Sup. (cat)' experiment uses \textit{supervised learning} to directly tune the multi-modal Gigapath baseline on both tasks on the train sets of the OOD datasets. 

\begin{table*}[]
\centering
\footnotesize
\begin{adjustbox}{width=0.72\linewidth}
\begin{tabular}{clccccc}
\hline
& & \textbf{BRCA}  & \textbf{GBMLGG} & \textbf{NSCLC} & \textbf{RCC} & \textbf{Overall} \\
\hline

\multicolumn{7}{c}{\textbf{Cancer Subtype Prediction}} \\
\hline
\parbox[t]{2mm}{\multirow{4}{*}{\rotatebox[origin=c]{90}{WSI}}} 
& \textbf{Gigapath LP} \cite{xu2024whole} & $0.612$ & $0.900$ & $0.821$ & $0.851$ & $0.796$\\
& \textbf{ABMIL} \cite{ilse2018attention} & $0.853 \pm 0.015$ & $0.931 \pm 0.040$ & $0.920 \pm 0.012$ & $0.921 \pm 0.017$ & $0.906$\\
& \textbf{TransMIL} \cite{shao2021transmil} & $0.828 \pm 0.011$ & $0.978 \pm 0.012$ & $0.934 \pm 0.007$ & $0.918 \pm 0.016$ & $0.915$\\
& \textbf{Gigapath (Tuned)} \cite{xu2024whole} & $0.860 \pm 0.013$ & $0.931 \pm 0.029$ & $0.916 \pm 0.012$ & $0.939 \pm 0.016$ & $0.912$\\
\hline
\parbox[t]{2mm}{\multirow{4}{*}{\rotatebox[origin=c]{90}{Genomics}}} & \textbf{MLP} & $0.752 \pm 0.032$ & $\underline{0.998} \pm 0.002$ & $0.926 \pm 0.007$ & $0.883 \pm 0.016$ & $0.890$\\
& \textbf{SNN} \cite{klambauer2017self} & $0.753 \pm 0.019$ & $0.991 \pm 0.002$ & $0.926 \pm 0.001$ & $0.902 \pm 0.008$ & $0.893$\\
& \textbf{S-MLP} \cite{elmarakeby2021biologically} & $0.839 \pm 0.015$ & $\mathbf{1.000} \pm 0.000$ & $0.941 \pm 0.005$ & $0.890 \pm 0.003$ & $0.917$\\
& \textbf{Gene Mixer} & $0.840 \pm 0.014$ & $\mathbf{1.000} \pm 0.000$ & $0.932 \pm 0.005$ & $0.898 \pm 0.036$ & $0.917$\\
\hline
\parbox[t]{2mm}{\multirow{9}{*}{\rotatebox[origin=c]{90}{Multi-modal}}} & \textbf{MCAT} \cite{Chen2021MultimodalImages} & $0.875 \pm 0.014$ & $0.973 \pm 0.005$ & $0.921 \pm 0.006$ & $0.933 \pm 0.028$ & $0.925$\\
& \textbf{SurvPath} \cite{jaume2024modeling} & $0.858 \pm 0.028$ & $0.995 \pm 0.007$ & $0.932 \pm 0.007$ & $0.936 \pm 0.030$ & $0.930$\\
& \textbf{ABMIL (cat)} \cite{ilse2018attention} & $0.861 \pm 0.010$ & $\mathbf{1.000} \pm 0.000$ & $0.951 \pm 0.003$ & $0.937 \pm 0.028$ & $0.937$\\
& \textbf{ABMIL (KP)} \cite{ilse2018attention} & $\underline{0.884} \pm 0.010$ & $\underline{0.998} \pm 0.002$ & $0.939 \pm 0.005$ & $0.945 \pm 0.031$ & $\underline{0.941}$\\
& \textbf{TransMIL (cat)} \cite{shao2021transmil} & $0.847 \pm 0.027$ & $\mathbf{1.000} \pm 0.000$ & $0.950 \pm 0.004$ & $\underline{0.957} \pm 0.020$ & $0.939$\\
& \textbf{TransMIL (KP)} \cite{shao2021transmil} & $0.868 \pm 0.017$ & $\mathbf{1.000} \pm 0.000$ & $0.946 \pm 0.005$ & $0.930 \pm 0.011$ & $0.936$\\
& \textbf{Gigapath (cat)} \cite{xu2024whole} & $0.850 \pm 0.017$ & $\underline{0.998} \pm 0.002$ & $0.924 \pm 0.011$ & $0.926 \pm 0.034$ & $0.925$\\
& \textbf{Gigapath (KP)} \cite{xu2024whole} & $0.821 \pm 0.020$ & $\mathbf{1.000} \pm 0.000$ & $\mathbf{0.963} \pm 0.003$ & $0.927 \pm 0.018$ & $0.928$\\
\cline{2-7}
& \textbf{ModalTune (Ours)} \rule{0pt}{2.4ex} &  $\mathbf{0.899} \pm 0.026$ &  $\mathbf{1.000} \pm 0.000$ &  $0.956 \pm 0.010$ &  $\mathbf{0.959} \pm 0.003$ &  $\mathbf{0.954}$\\
& \textbf{ModalTune Pan-Cancer (Ours)} \rule[-0.9ex]{0pt}{0pt} & $0.858 \pm 0.001$ & $0.990 \pm 0.009$ & $\underline{0.958} \pm 0.004$ & $0.902 \pm 0.033$ & $0.927$\\

\hline

\multicolumn{7}{c}{\textbf{Survival Prediction}} \\
\hline
\parbox[t]{2mm}{\multirow{4}{*}{\rotatebox[origin=c]{90}{WSI}}} 
& \textbf{Gigapath LP} \cite{xu2024whole} & $0.647$ & $0.795$ & $0.562$ & $0.679$ & $0.671$\\
& \textbf{ABMIL} \cite{ilse2018attention} & $0.712 \pm 0.004$ & $0.854 \pm 0.004$ & $0.582 \pm 0.002$ & $0.670 \pm 0.004$ & $0.704$\\
& \textbf{TransMIL} \cite{shao2021transmil} & $0.742 \pm 0.015$ & $0.868 \pm 0.009$ & $0.586 \pm 0.011$ & $0.676 \pm 0.003$ & $0.718$\\
& \textbf{Gigapath (Tuned)}  \cite{xu2024whole} & $0.680 \pm 0.024$ & $0.824 \pm 0.017$ & $0.546 \pm 0.005$ & $0.685 \pm 0.012$ & $0.684$\\
\hline
\parbox[t]{2mm}{\multirow{4}{*}{\rotatebox[origin=c]{90}{Genomics}}} & \textbf{MLP} & $0.629 \pm 0.039$ & $0.884 \pm 0.004$ & $0.542 \pm 0.013$ & $0.720 \pm 0.002$ & $0.694$\\
& \textbf{SNN} \cite{klambauer2017self} & $0.519 \pm 0.051$ & $0.883 \pm 0.010$ & $0.508 \pm 0.012$ & $0.723 \pm 0.008$ & $0.659$\\
& \textbf{S-MLP} \cite{elmarakeby2021biologically} & $0.749 \pm 0.030$ & $0.887 \pm 0.002$ & $0.571 \pm 0.011$ & $\underline{0.735} \pm 0.003$ & $\underline{0.736}$\\
& \textbf{Gene Mixer} & $\underline{0.762} \pm 0.049$ & $0.870 \pm 0.007$ & $0.556 \pm 0.033$ & $0.690 \pm 0.009$ & $0.719$\\
\hline
\parbox[t]{2mm}{\multirow{9}{*}{\rotatebox[origin=c]{90}{Multi-modal}}} & \textbf{MCAT} \cite{Chen2021MultimodalImages} & $0.673 \pm 0.044$ & $0.880 \pm 0.006$ & $0.592 \pm 0.002$ & $0.697 \pm 0.004$ & $0.710$\\
& \textbf{SurvPath} \cite{jaume2024modeling} & $0.741 \pm 0.031$ & $\underline{0.895} \pm 0.003$ & $\mathbf{0.613} \pm 0.010$ & $0.677 \pm 0.011$ & $0.732$\\
& \textbf{ABMIL (cat)} \cite{ilse2018attention} & $0.736 \pm 0.007$ & $\mathbf{0.896} \pm 0.004$ & $0.605 \pm 0.011$ & $0.690 \pm 0.004$ & $0.732$\\
& \textbf{ABMIL (KP)} \cite{ilse2018attention} & $0.685 \pm 0.008$ & $0.895 \pm 0.002$ & $0.594 \pm 0.010$ & $0.699 \pm 0.003$ & $0.718$\\
& \textbf{TransMIL (cat)} \cite{shao2021transmil} & $0.666 \pm 0.018$ & $0.872 \pm 0.030$ & $0.595 \pm 0.003$ & $0.689 \pm 0.004$ & $0.705$\\
& \textbf{TransMIL (KP)} \cite{shao2021transmil} & $0.689 \pm 0.014$ & $0.889 \pm 0.003$ & $0.541 \pm 0.023$ & $0.713 \pm 0.003$ & $0.708$\\
& \textbf{Gigapath (cat)} \cite{xu2024whole} & $0.678 \pm 0.004$ & $0.861 \pm 0.023$ & $0.573 \pm 0.003$ & $0.693 \pm 0.016$ & $0.701$\\
& \textbf{Gigapath (KP)} \cite{xu2024whole} & $0.650 \pm 0.043$ & $0.865 \pm 0.032$ & $0.563 \pm 0.005$ & $0.686 \pm 0.016$ & $0.691$\\
\cline{2-7}
& \textbf{ModalTune (Ours)} \rule{0pt}{2.4ex} & $\mathbf{0.772} \pm 0.008$ & $0.879 \pm 0.004$ & $\underline{0.608} \pm 0.023$ & $\mathbf{0.743} \pm 0.004$ & $\mathbf{0.750}$\\
& \textbf{ModalTune Pan-Cancer (Ours)} \rule[-0.9ex]{0pt}{0pt} & $0.757 \pm 0.039$ & $0.860 \pm 0.006$ & $0.586 \pm 0.020$ & $0.705 \pm 0.007$ & $0.727$\\

\hline
\end{tabular}
\end{adjustbox}
\caption{Cancer subtype prediction balanced accuracy and survival prediction C-index scores across 4 cancer types. Best model in \textbf{bold}, second best is \underline{underlined}. Here, LP refers to linear probing, cat refers to concatenation, and KP refers to Kronecker product.}
\label{tab: diagnosis_and_survival}
\end{table*}

\begin{figure*}[t]
  \centering
   \includegraphics[width=0.71\linewidth]{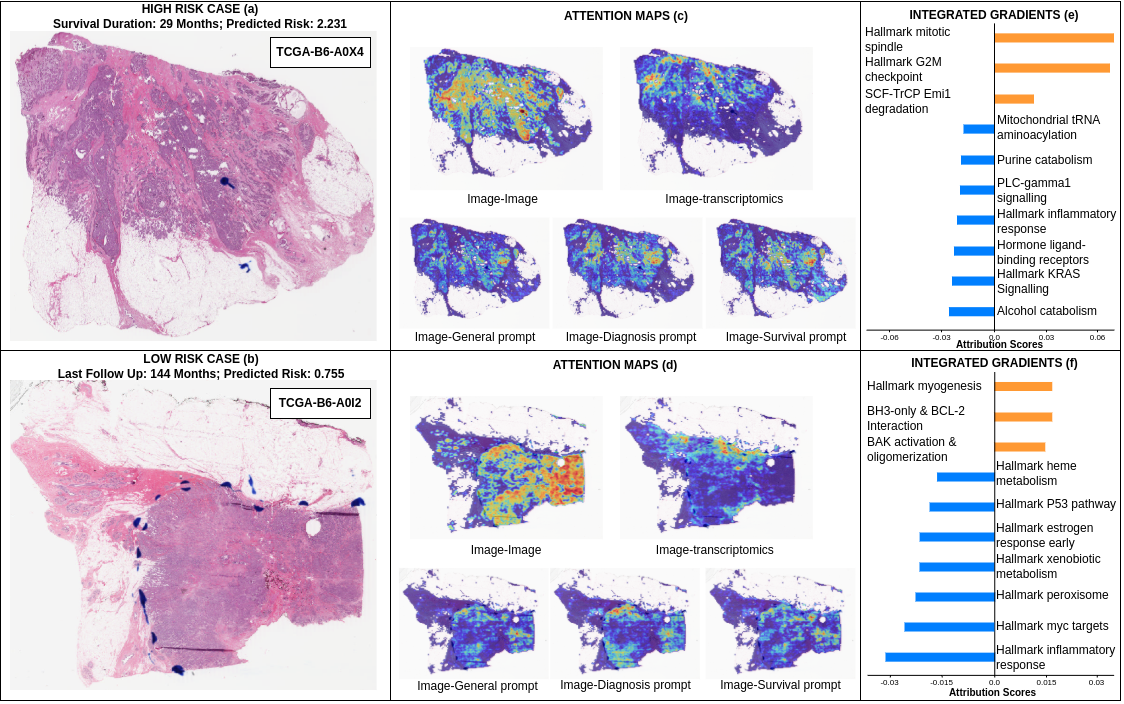}
   \caption{Qualitative analysis of breast cancer cases, highlighting a high-risk (\textbf{a}) and a low-risk case (\textbf{b}). (1) \textbf{Attention Maps (c,d)} depict cross-modal and cross-task interactions, with heatmap colors indicating importance (red: high, blue: low). (2) \textbf{Integrated Gradients (e,f)} shows the top 10 pathways influencing risk, with orange bars indicating pathways increased risk and blue bars indicating pathways decreased risk. The analysis highlights biologically relevant pathways (e.g., G2M checkpoint) and morphologically significant regions.}
   \label{fig: interpretability}
\end{figure*}

\subsection{Quantitative Analysis}
\label{sec:quant_analysis}

To address class imbalances in subtype classification, we report balanced accuracy, while survival prediction is evaluated using C-indices. \cref{tab: diagnosis_and_survival} presents subtype classification and survival prediction results, showing the mean and standard deviation across three random seed runs. We report ablation studies on key design choices, including Modal Adapters, text embeddings, text encoders, task prompts, and Projectors in the Supp. \cref{sec:supp_ablations}.

\noindent \textbf{Overall Results:} On average, we found ModalTune surpassed all other baselines on subtype classification and survival prediction tasks on multiple cancer sites, with $+1.4\%$ and $+1.9\%$ overall improvement, respectively, over the best baseline. ModalTune also greatly improved over the standard image-only SLFM framework, Gigapath (Tuned) ($+4.6\%$ in subtype, $+9.6\%$ in survival prediction). These large improvements can be partially attributed to the introduction of the additional bulk transcriptomics modality. However, a fully-tuned Gigapath model with a late fusion of transcriptomics features (Gigapath (cat) and (KP)) still performs $2.8\%$ worse on classification and $7.0\%$ worse on survival prediction compared to ModalTune, indicating the benefits of our overall framework. Though TITAN is a stronger standalone model, applying ModalTune with it yielded similar trends (supp. \cref{sec:supp_titan}).

\noindent \textbf{Importance of Robust Multi-modal Interactions}: 
The best modality for prediction varied by task and cancer site. For instance, WSI-baselines outperformed Genomics-baselines in BRCA subtyping, while the reverse was observed for RCC survival prediction. Combining modalities generally enhanced performance, as seen in GBMLGG, where adding bulk transcriptomics led to near-perfect subtyping in multi-modal models, with a +7.4\% improvement from single to multi-modal Gigapath. Further investigation using a linear classifier on transcriptomics (test balanced accuracy: 0.99) identified four genes—CDC26, IDH1, IL13RA2, and IL8—explaining 97.5\% of accuracy, with IDH1 being a well-documented biomarker \cite{verhaak2010integrated}, potentially trivializing classification. Thus, multi-modal results on GBMLGG serve as a ``sanity check'' for the transcriptomics signal, which fusion models—except MCAT—effectively retained. However, naïve fusion of multi-modal features was not always effective, especially for survival prediction, where the best multi-modal TransMIL model (KP) dropped 1.5\% compared to Gene Mixer, and Gigapath (cat) dropped 2.5\%. These findings emphasize the need for a \textit{robust mixing strategy} when integrating multi-modal features.

\noindent \textbf{Generalization Study}: 
As shown in \cref{tab: generalizability}, the fully tuned Gigapath (cat) models severely overfit on BRCA, resulting in poor performance on both subtype and survival prediction tasks. These models fail to extract meaningful features for subtyping, defaulting to predicting the most common class (balanced accuracy of $\sim$0.5). We also observe that ModalTune consistently outperforms Gigapath LP across both tasks and cancer sites. Furthermore, ModalTune Pan-Cancer demonstrated superior generalizability, performing only 2.4\% worse than a \textit{fully-supervised} Gigapath (cat) network in subtype prediction and actually outperforming it by 1.2\% in survival prediction. ModalTune's ability to extract highly relevant features without exposure to OOD datasets during fine-tuning underscores its capacity to maintain the generalizability of SLFMs. This result is particularly promising for the downstream application of ModalTune to smaller unseen datasets. 

\begin{table}[]
\centering
\small
\begin{adjustbox}{width=0.76\linewidth}
\begin{tabular}{lccccc}
\hline
 & \textbf{COADREAD} & \textbf{BLCA} \\
\hline
\multicolumn{3}{c}{\textbf{Cancer Subtype Prediction}} \\
\hline
\textbf{Gigapath LP} & $0.510$ & $0.569$ \\
\textbf{Gigapath Sup. (cat)} & $ 0.581\pm0.006 $ & $ 0.703 \pm 0.018 $ \\
\hline
\textbf{Gigapath Cls. (cat)} & 
$0.504 \pm 0.030$ & $0.497 \pm 0.005$ \\
\textbf{Gigapath Surv. (cat)} & $0.500 \pm 0.000$ & $0.497 \pm 0.005$ \\
\textbf{ModalTune} & $\textbf{0.574} \pm 0.024$ & $\underline{0.664} \pm 0.025$ \\
\textbf{ModalTune Pan-Cancer} & $\underline{0.564} \pm 0.034$ & $\textbf{0.689} \pm 0.035$\\
\hline
\multicolumn{3}{c}{\textbf{Survival Prediction}} \\
\hline
\textbf{Gigapath LP} & $0.482$ & $0.603$ \\
\textbf{Gigapath Sup. (cat)} & $ 	0.528\pm0.023 $ & $ 0.673\pm0.020 $ \\
\hline
\textbf{Gigapath Cls. (cat)} & $0.479 \pm 0.042$ & $0.610 \pm 0.063$ \\
\textbf{Gigapath Surv. (cat)} & $0.512 \pm 0.061$ & $0.552 \pm 0.052$ \\
\textbf{ModalTune} & $\underline{0.539} \pm 0.068$ & $\underline{0.629} \pm 0.041$ \\
\textbf{ModalTune Pan-Cancer} &  $\textbf{0.543} \pm 0.020$ & $\textbf{0.672} \pm 0.046$ \\
\hline
\end{tabular}
\end{adjustbox}
\caption{Generalization study on OOD datasets using different models, compared with Gigapath Sup. (cat) trained directly on the OOD data. Best OOD model in \textbf{bold}, second best is \underline{underlined}.}
\label{tab: generalizability}
\end{table}

\noindent \textbf{Pan-Cancer Results:} 
As shown in \cref{tab: diagnosis_and_survival}, training a single ModalTune Pan-Cancer model by merging cancer sites did not outperform training models independently for each site on in-domain datasets, with a decrease of 2.8\% in classification and 3.0\% in survival prediction. This aligns with findings in \cite{mahmoodl9:online}, where a specialized model outperformed the pan-cancer model for molecular status prediction. However, the Pan-Cancer model is significantly more resource-efficient, requiring only one model instead of one per cancer site. As the number of sites $N_{sites}$ and tasks $T$ increase, scalability and efficiency benefits become more pronounced, surpassing simple baselines. This highlights a trade-off where the Pan-Cancer model enhances resource efficiency with minor performance compromises.

We hypothesize the performance drop on in-domain datasets is due to variations in task difficulty across cancer sites. Training standard ModalTune revealed variability in the optimal validation epoch per site—NSCLC reached peak performance 10 epochs earlier than GBMLGG, which required more training to converge. In the pan-cancer model, early stopping optimized for NSCLC led to underfitting for GBMLGG, while extended training overfitted NSCLC. Despite these convergence challenges reducing in-domain performance, exposure to diverse cancer sites in the pan-cancer setup greatly enhances generalizability on OOD datasets (\cref{tab: generalizability}). Thus, we believe the pan-cancer approach is promising and emphasize the need for optimization strategies to recover in-domain performance.

\subsection{Qualitative Analysis} 
To assess pathway contributions and modality interactions, we analyze two breast cancer cases from high and low-risk categories, as shown in \cref{fig: interpretability}. See Supp. for more qualitative analysis of t-SNE and Kaplan-Meier (\cref{sec:supp_qualanalysis}).

\noindent \textbf{Pathway Contributions}: We applied Integrated Gradients \cite{sundararajan2017axiomatic} to analyze pathway contributions as shown in \cref{fig: interpretability} (e,f). In the high-risk case, incorporating pathway signals raised the risk score from 2.07 to 2.23. Key contributors included mitotic spindle assembly \cite{walczak2008mechanisms}, G2M checkpoint regulation \cite{lobrich2007impact}, and SCF-beta-TrCP Mediated Emi1 degradation \cite{margottin2003prophase}, consistent with their roles in cell cycle progression and tumor proliferation. In both cases, the inflammatory response \cite{zhao2022novel} lowered risk, reflecting its association with reduced oncogenic activity. Metabolic pathways like alcohol and purine catabolism \cite{watts1974molecular} also reduced risk but require further study. In the low-risk case, the score dropped from 1.09 to 0.75, driven by inflammatory response \cite{zhao2022novel}, p53 activation \cite{gasco2002p53}, MYC targets \cite{schulze2020myc}, and estrogen signaling suppression \cite{oshi2020degree}—aligning with established roles in modulating tumor progression or suppression. Interestingly, myogenesis \cite{krauss2005close} and apoptosis regulation (BH3-only \& BCL-2 interaction, BAK activation, and olgomerization) contributed to increased risk, with myogenesis also linked to risk in SurvPath’s analysis \cite{jaume2024modeling}, warranting further study.

\noindent \textbf{Interaction Analysis}: We analyzed cross- and self- attention maps to understand different learned interactions, as shown in \cref{fig: interpretability} (c,d). Specifically, we examined the self-attention of LongNet \texttt{[CLS]} tokens with patch tokens (image-image), averaged cross-attention of pathway tokens with patch tokens (image-transcriptomics), and cross-attention of patch tokens with task prompts (image-\texttt{[task]} prompt). Our findings show that image-image self-attention primarily focuses on tumor regions, capturing key pathological features, while image-transcriptomics attention highlights tumor-associated stroma. Differences in attention maps for subtype prediction and survival task prompts suggest both shared and distinct feature utilization, with general prompts integrating elements from both.

\section{Conclusion}
In this study, we introduced ModalTune, a flexible fine-tuning framework that maximizes dataset usage for SLFMs in histopathology. To our knowledge, ModalTune is the first multi-modal, multi-task \textit{and} pan-cancer tuning framework. By leveraging inter-modal and inter-task interactions, ModalTune outperformed all single-task baselines while retaining generalizability. Its resource-efficient variant, ModalTune Pan-Cancer, further improves generalization on small OOD datasets, outperforming other fine-tuned models by a large margin while remaining competitive with supervised baselines. This underscores ModalTune's ability to enhance SLFMs' capacity for handling smaller datasets.

Despite its strengths, ModalTune faces two key challenges. Firstly, because the interaction between multi-modal and image features is mediated by cross-attention, it can be computationally expensive. We mitigated this by compressing pathway tokens, but future work may explore cheaper formulations of cross-attention. Secondly, the model may underfit one task while overfitting another due to varying task complexities, as seen in our pan-cancer experiments. Future work will explore strategies to mitigate the pan-cancer convergence issue, and ModalTune's adaptability to other SLFMs. Looking ahead, we also aim to incorporate additional tasks and modalities to ModalTune.

\section{Acknowledegments}
This work was supported by funding from the Canadian Institutes of Health Research (CIHR grant \#162327), Canada Foundation for Innovation (40206), and the Ontario Research Fund. T.X. is supported by the NSERC PGS-D award and Google PhD Fellowship. A.L.M. is partially supported by the Tory Family Chair in Oncology. 

{
    \small
    \bibliographystyle{ieeenat_fullname}
    \bibliography{main}
}

\clearpage
\setcounter{page}{1}
\maketitlesupplementary

In this Supplementary Material, we have provided additional details for the following:
\begin{itemize}
    \item Hyperparameters (\cref{sec:supp_Hyperparameters})
    \item Class Groupings (\cref{sec:supp_classgroupings})
    \item Text Construction (\cref{sec:supp_Text})
    \item Text Embedding Analysis (\cref{sec:supp_textembeddinganalysis})
    \item Ablations (\cref{sec:supp_ablations})
    \item Experiments with TITAN (\cref{sec:supp_titan})
    \item Additional Modalities (\cref{sec:supp_multimodal})
    \item Qualitative Analysis (\cref{sec:supp_qualanalysis})
    
\end{itemize}

\section{Hyperparameters}
\label{sec:supp_Hyperparameters}

We display additional hyperparameters used to train ModalTune in \cref{tab: hyperparams}. 

\section{Class Groupings}
\label{sec:supp_classgroupings}
In the TCGA dataset, clinician-annotated cancer subtypes are highly detailed. However, predicting each individual subtype is infeasible due to their large number and the limited cases for certain subtypes. Therefore, we grouped the subtypes into broader categories and \textit{Rare-set class} with the help of OncoTree code \cite{kundra2021oncotree}, as shown in \cref{tab:class_groupings}. Subtypes in \textit{Rare-set classes} were used for generating text embeddings but were excluded from baseline training due to their low sample size. Additionally, we merged subtypes with minor differences (e.g., "Squamous cell carcinoma, keratinizing" and "Squamous cell carcinoma, large cell, nonkeratinizing" were grouped as "Squamous cell carcinoma"). The final subtype bins for different cancer types are presented in \cref{tab:class_groupings}. 

For survival prediction task, we categorized survival durations into four bins, ensuring an approximately equal number of patients in each. The bin limits were then used to generate textual descriptions, as illustrated in \cref{tab:surv_groupings}.
\begin{table}[t]
\begin{adjustbox}{width=\linewidth}
\centering
\begin{tabular}{|l|p{10cm}|}
\hline
\textbf{Cancer Type} & \textbf{Class Groupings} \\ \hline
\textbf{BRCA}   & 0: Infiltrating duct carcinoma\\ 
                & 1: Lobular carcinoma\\ 
                & Rare-set: infiltrating duct and lobular carcinoma, infiltrating (duct/lobular) mixed with other types of carcinoma, mucinous adenocarcinoma, metaplastic carcinoma, medullary carcinoma, intraductal papillary adenocarcinoma with invasion, tubular adenocarcinoma, adenoid cystic carcinoma, cribriform carcinoma \\ \hline
\textbf{GBMLGG} & 0: Glioblastoma\\ 
                & 1: Mixed glioma, Oligodendroglioma, Astrocytoma, Oligodendroglioma anaplastic, Astrocytoma anaplastic\\ 
                \hline
\textbf{NSCLC}  & 0: Lung adenocarcinoma\\ 
                & 1: Lung squamous cell carcinoma\\ 
                & Rare-set: lung bronchiolo-alveolar carcinoma, lung papillary adenocarcinoma, lung acinar cell carcinoma, lung basaloid squamous cell carcinoma, lung solid carcinoma, lung signet ring cell carcinoma, lung papillary squamous cell carcinoma, lung micropapillary carcinoma \\ \hline
\textbf{RCC}    & 0: Papillary renal cell carcinoma\\ 
                & 1: Renal clear cell carcinoma\\ 
                & 2: Chromophobe renal cell carcinoma\\ 
                 \hline
\textbf{COADREAD}    & 0: Colon adenocarcinoma\\ 
                & 1: Rectal adenocarcinoma\\ 
                & Rare-set: colon mucinous adenocarcinoma, rectal mucinous adenocarcinoma, rectal adenocarcinoma in tubolovillous adenoma, rectal tubular adenocarcinoma, colon papillary adenocarcinoma   \\ 
                 \hline
\textbf{BLCA}    & 0: Transitional cell carcinoma\\ 
                & 1: Papillary transitional cell carcinoma\\ 
                 \hline
\end{tabular}
\end{adjustbox}
\caption{Cancer subtype groupings for different cancer types}
\label{tab:class_groupings}
\end{table}

\begin{table}[t]
\begin{adjustbox}{width=\linewidth}
\centering
\begin{tabular}{|l|p{10cm}|}
\hline
\textbf{Cancer Type} & \textbf{Duration Bins} \\ \hline
\textbf{BRCA}   & 0: before 15 months \\ 
                & 1: between 15 and 27 months\\ 
                & 2: between 27 and 55 months\\
                & 3: between 55 and 283 months\\ \hline
\textbf{GBMLGG}   & 0: before 8 months\\ 
                & 1: between 8 and 17 months \\ 
                & 2: between 17 and 31 months \\
                & 3: between 31 and 211 months \\ \hline
\textbf{NSCLC}   & 0: before 12 months \\ 
                & 1: between 12 and 22 months \\ 
                & 2: between 22 and 39 months \\
                & 3: between 39 and 238 months \\ \hline
\textbf{RCC}   & 0: before 16 months\\ 
                & 1: between 16 and 34 months \\ 
                & 2: between 34 and 62 months \\
                & 3: between 62 and 169 months \\ \hline
\textbf{COADREAD}   & 0: before 12 months\\ 
                & 1: between 12 and 21 months \\ 
                & 2: between 21 and 36 months \\
                & 3: between 36 and 148 months \\ \hline
\textbf{BLCA}   & 0: before 11 months\\ 
                & 1: between 11 and 18 months \\ 
                & 2: between 18 and 30 months \\
                & 3: between 30 and 163 months \\ \hline
\end{tabular}
\end{adjustbox}
\caption{Duration bins for different cancer types. The text is used by ModalTune and the bin labels are used as labels by baselines}
\label{tab:surv_groupings}
\end{table}

\begin{table}[t]
\centering
\small
\begin{adjustbox}{width=\linewidth}
\begin{tabular}{lc}
\hline
\textbf{Parameter} & \textbf{Value} \\
\hline
\multicolumn{2}{c}{\textbf{Slide encoder settings}} \\
\hline
Embedding dim, $D$ & 768 \\
Layers, $L$ & 12 \\
Attention heads & 16 \\
Feedfoward dim & 3072 \\
Dilated attention segment lengths & $[1024, 2048, 4096, 8192, 16384]$ \\
Dilated attention ratios & $[1, 2, 4, 8, 16]$ \\
Activation function & GELU \\
\hline
\multicolumn{2}{c}{\textbf{Transcriptomics encoder settings}} \\
\hline
Embedding dim, $D_{gp}$ & 256 \\
\# Compressed pathways, $N_t$ & 64 \\
Layers & 3 \\
Feedforward expansion ratio & 0.5 \\
Dropout & 0.25 \\
Activation function & GELU \\
\hline
\multicolumn{2}{c}{\textbf{ModalTune settings}} \\
\hline
Embedding dim, $D$ & 768 \\
Adapter blocks, $B$ & 3 \\
Adapter cross-attention heads & 12 \\
Adapter feedforward expansion ratio & 0.25 \\
Adapter dropout & 0.1 \\
Adapter initial gamma, $\gamma^i_0$ & 0 \\
Final output dim, $D_{final}$ & 256 \\
Text embedding dim, $D_{text}$ & 512 \\
\hline
\multicolumn{2}{c}{\textbf{Training settings}} \\
\hline
Epochs & 30 \\
Optimizer & AdamW \\
Max LR & 1e-4 \\
LR scheduler & LinearWarmupCosineAnnealing \\
Warmup epochs & 10 \\
Weight decay & 0.0005 \\
Batch size & 1 \\
\hline
\multicolumn{2}{c}{\textbf{Inference settings}} \\
\hline
Logistic regression max iters & 200 \\
Logistic regression solver & liblinear \\
CPH penalizer & 0.1 \\
\hline
\end{tabular}
\end{adjustbox}
\caption{Additional ModalTune hyperparameters.}
\label{tab: hyperparams}
\end{table}

\section{Text Construction}
\label{sec:supp_Text}

We construct the embedded text prompts used to train ModalTune from clinical tables in \texttt{.csv} format available with all TCGA slides. This is done by firstly cleaning the table entries and converting them to natural language to take better advantage of semantic relationships in text. For example, we convert node status 0 (N0) to: "cancer has not spread to lymph nodes". For text related to tumor, node, metastastasis (TNM) staging, we also bin sub-categories of stages into a single stage to reduce variability of text embeddings (e.g. T1a, T1b, and T1c become "tumor stage 1"). We use the cancer subtype texts obtained after the pre-processing steps described in  (\cref{sec:supp_classgroupings}). 

 We then describe the type of event (censored or an event occurred) along with a description of the bin as shown in \cref{tab:surv_groupings}. For example, a patient censored at 144 months would have the status: "The patient was censored between 55 and 283 months". 

\begin{figure}[t]
  \centering
   \includegraphics[width=1.0\linewidth]{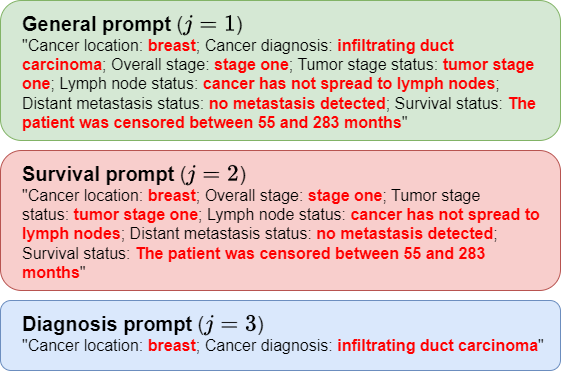}
   \caption{Example text prompts generated for the low-risk patient in \cref{fig: interpretability}. Text in bold and red are directly obtained from clinical tables in TCGA after being clearned and converted to natural language.}
   \label{fig:text_example}
\end{figure}

Task-specific text prompts contain information that is directly relevant to the task. For subtype classification ($j=3$), we included the cancer site and the cancer subtype. For survival prediction ($j=2$), we included the cancer site, TNM stages, and the survival status of the patient as described above. For the general task ($j=1$), we merged the two prompts (only mentioning cancer site a single time). We chose to include TNM staging information for the survival and general tasks to better estimate and \textit{delineate} risk between patients. We found staging to be prognostic, improving performance over solely relying on survival duration bins (\cref{tab: textembeddingperformance}). In any cases where TNM stage is not available (like the full patient cohort of GBMLGG), we simply omit stage-related text from the prompt. Example text generated for the low-risk patient in \cref{fig: interpretability} is displayed in \cref{fig:text_example}.

\section{Text Embedding Analysis}
\label{sec:supp_textembeddinganalysis}
We analyze performance across different tasks using ($j=1$) general task embeddings to confirm whether the text embeddings capture task-relevant information. Logistic regression is used for cancer subtyping and duration bin prediction, while cox proportional hazards model is applied for survival prediction. We report the mean and standard deviation of balanced accuracy for classification tasks and the C-index for survival prediction, averaged over three random seeds. We used the same splits as those used for other experiments. Overall, as shown in \cref{tab: textembeddingperformance}, we observe near-perfect performance across all tasks. Adding stage information to the general embedding slightly improves the C-index. Overall, as shown in \cref{tab: textembeddingperformance}, we observe near-perfect performance across all tasks. Adding stage information to the general embedding slightly improves the C-index. Random projections preserve the tight clusters, maintaining performance across multiple seeds and exhibiting similar performance in all of the tasks with minor degradations. This does not impact ModalTune; in fact, it enhances its performance in both cancer subtype prediction and survival prediction, as shown in \cref{tab: ablations}.
\begin{table*}[]
\centering
\begin{tabular}{lcc}
\hline
\textbf{Tasks}               & \textbf{Text Embedding} & \textbf{Text Embedding after random projection} \\ \hline
\textbf{Cancer Subtyping}    &     $1.000 \pm 0.000$           &  $1.000 \pm 0.000$                                       \\
\textbf{Duration Bins }      &      $1.000 \pm 0.000$           &        $0.998 \pm 0.002$                                  \\
\textbf{Survival prediction w/o stage} & $0.966 \pm 0.000$&  $0.968 \pm 0.000$\\  
\textbf{Survival prediction} &       $0.972 \pm 0.000$         &        $0.970 \pm 0.001$                                 \\ \hline
\end{tabular}
\caption{Text embedding performance on different tasks on TCGA BRCA. We report balanced accuracy for cancer subtyping (2 class classification) and duration bins (4 class classification), and C-index for Survival prediction tasks}
\label{tab: textembeddingperformance}
\end{table*}

\begin{table*}[t]
\centering
\small
\begin{adjustbox}{width=0.8\linewidth}
\begin{tabular}{lccccc}
\hline
\textbf{Ablation} & \textbf{BRCA} & \textbf{GBMLGG} & \textbf{NSCLC} & \textbf{RCC} & \textbf{Overall} \\
\hline
\multicolumn{6}{c}{\textbf{Cancer Subtype Prediction}} \\
\hline
\textbf{Single Modal} & $0.887 \pm 0.029$ & $0.937 \pm 0.012$ & $0.926 \pm 0.002$ & $0.930 \pm 0.009$ & $0.920$ \\
\textbf{No Text Embedding} & $0.885 \pm 0.012$ & $\underline{0.998} \pm 0.002$ & $\underline{0.956} \pm 0.002$ & $\underline{0.954} \pm 0.005$ & $\underline{0.948}$ \\
\textbf{Single Task Prompt} & $0.855 \pm 0.005$ & $0.995 \pm 0.004$ & $0.950 \pm 0.005$ & $0.939 \pm 0.016$ & $0.935$ \\
\textbf{ABMIL (cat) w/ text emb.} & $0.874 \pm 0.024$ & $0.973 \pm 0.021$ & $0.934 \pm 0.006$ & $0.906 \pm 0.012$ & $0.922$ \\
\textbf{ModalTune w/ LoRA} & $0.883 \pm 0.026$ & $\underline{0.998} \pm 0.003$ & $\textbf{0.960} \pm 0.001$ & $0.920 \pm 0.033$ & $0.940$ \\
\textbf{ModalTune w/ MedLlama v3.1} & $0.853 \pm 0.011$ & $0.993 \pm 0.006$ & $0.919 \pm 0.009$ & $0.918 \pm 0.013$ & $0.921$ \\
\textbf{No Projector} & $0.891 \pm 0.024$ & $0.997 \pm 0.002$ & $0.948 \pm 0.005$ & $0.951 \pm 0.011$ & $0.947$ \\
\textbf{Trainable Projector} & $0.612 \pm 0.017$ & $0.542 \pm 0.023$ & $0.768 \pm 0.024$ & $0.685 \pm 0.078$ & $0.652$ \\
\textbf{Model-side Projector} & $\underline{0.898} \pm 0.003$ & $0.993 \pm 0.003$ & $\underline{0.956} \pm 0.006$ & $0.918 \pm 0.044$ & $0.941$ \\
\textbf{ModalTune} &  $\textbf{0.899} \pm 0.026$ &  $\textbf{1.000} \pm 0.000$ &  $\underline{0.956} \pm 0.010$ &  $\textbf{0.959} \pm 0.003$ &  $\textbf{0.954}$\\
\hline
\multicolumn{6}{c}{\textbf{Survival Prediction}} \\
\hline
\textbf{Single Modal} & $0.730 \pm 0.025$ & $0.821 \pm 0.016$ & $0.586 \pm 0.013$ & $0.689 \pm 0.018$ & $0.707$ \\
\textbf{No Text Embedding} & $0.724 \pm 0.024$ & $0.881 \pm 0.006$ & $\textbf{0.631} \pm 0.010$ & $0.682 \pm 0.022$ & $0.730$ \\
\textbf{Single Task Prompt} & $0.757 \pm 0.014$ & $0.872 \pm 0.005$ & $0.585 \pm 0.018$ & $\underline{0.741} \pm 0.007$ & $0.739$ \\
\textbf{ABMIL (cat) w/ text emb.} & $0.742 \pm 0.016$ & $0.869 \pm 0.024$ & $0.603 \pm 0.033$ & $0.710 \pm 0.011$ & $0.731$ \\
\textbf{ModalTune w/ LoRA} & $0.756 \pm 0.038$ & $\textbf{0.894} \pm 0.008$ & $0.598 \pm 0.011$ & $0.728 \pm 0.032$ & $\underline{0.744}$ \\
\textbf{ModalTune w/ MedLlama v3.1} & $0.752 \pm 0.032$ & $0.868 \pm 0.011$ & $0.603 \pm 0.036$ & $0.733 \pm 0.012$ & $0.739$ \\
\textbf{No Projector} & $0.726 \pm 0.007$ & $0.868 \pm 0.005$ & $\underline{0.612} \pm 0.028$ & $0.714 \pm 0.016$ & $0.730$ \\
\textbf{Trainable Projector} & $0.693 \pm 0.029$ & $0.803 \pm 0.016$ & $0.610 \pm 0.008$ & $0.694 \pm 0.027$ & $0.700$ \\
\textbf{Model-side Projector} & $\underline{0.771} \pm 0.037$ & $\underline{0.888} \pm 0.007$ & $0.594 \pm 0.009$ & $0.712 \pm 0.019$ & $0.742$ \\
\textbf{ModalTune} & $\textbf{0.772} \pm 0.008$ & $0.879 \pm 0.004$ & $0.608 \pm 0.023$ & $\textbf{0.743} \pm 0.004$ & $\textbf{0.750}$\\
\hline
\end{tabular}
\end{adjustbox}
\caption{Ablations across different tasks and cancer types investigating key design choices of ModalTune. Best model in \textbf{bold}, second best is \underline{underlined}}
\label{tab: ablations}
\end{table*}

\begin{table*}[t]
\centering
\small
\begin{adjustbox}{width=0.8\linewidth}
\begin{tabular}{lccccc}
\hline
\textbf{TITAN exp.} & \textbf{BRCA} & \textbf{GBMLGG} & \textbf{NSCLC} & \textbf{RCC} & \textbf{Overall} \\
\hline
\multicolumn{6}{c}{\textbf{Cancer Subtype Prediction}} \\
\hline
\textbf{TITAN LP} \cite{Ding2024MultimodalPathology} & $0.809$ & $0.965$ & $0.941$ & $0.943$ & $0.914$ \\
\textbf{TITAN (Tuned)} \cite{Ding2024MultimodalPathology} & $0.845 \pm 0.007$ & $0.948 \pm 0.020$ & $0.938 \pm 0.002$ & $\textbf{0.951} \pm 0.003$ & $0.920$ \\
\textbf{TITAN (cat)} \cite{Ding2024MultimodalPathology} & $\underline{0.849} \pm 0.010$ & $\textbf{0.998} \pm 0.002$ & $0.940 \pm 0.018$ & $0.941 \pm 0.016$ & $\underline{0.932}$ \\
\textbf{TITAN (KP)} \cite{Ding2024MultimodalPathology} & $0.825 \pm 0.011$ & $\textbf{0.998} \pm 0.003$ & $\textbf{0.955} \pm 0.003$ & $\underline{0.949} \pm 0.022$ & $\underline{0.932}$ \\
\textbf{ModalTune TITAN} &  $\textbf{0.872} \pm 0.013$ &  $\underline{0.997} \pm 0.002$ &  $\underline{0.950} \pm 0.003$ &  $0.948 \pm 0.012$ &  $\textbf{0.942}$\\
\hline
\multicolumn{6}{c}{\textbf{Survival Prediction}} \\
\hline
\textbf{TITAN LP} \cite{Ding2024MultimodalPathology} & $0.710$ & $0.770$ & $0.552$ & $0.677$ & $0.677$ \\
\textbf{TITAN (Tuned)} \cite{Ding2024MultimodalPathology} & $0.732 \pm 0.009$ & $0.832 \pm 0.006$ & $\textbf{0.620} \pm 0.005$ & $0.717 \pm 0.002$ & $0.725$ \\
\textbf{TITAN (cat)} \cite{Ding2024MultimodalPathology} & $\underline{0.745} \pm 0.046$ & $0.850 \pm 0.010$ & $\underline{0.604} \pm 0.008$ & $\textbf{0.729} \pm 0.008$ & $\underline{0.732}$ \\
\textbf{TITAN (KP)} \cite{Ding2024MultimodalPathology} & $0.739 \pm 0.018$ & $\textbf{0.866} \pm 0.018$ & $0.571 \pm 0.008$ & $0.719 \pm 0.008$ & $0.724$ \\
\textbf{ModalTune TITAN} &  $\textbf{0.753} \pm 0.012$ &  $\underline{0.858} \pm 0.016$ &  $\underline{0.604} \pm 0.012$ &  $\underline{0.725} \pm 0.023$ &  $\textbf{0.735}$\\
\hline
\end{tabular}
\end{adjustbox}
\caption{Cancer subtype prediction balanced accuracy and survival prediction C-index scores across 4 cancer types for TITAN slide encoder. Best model in \textbf{bold}, second best is \underline{underlined}. Here, LP refers to linear probing, cat refers to concatenation, and KP refers to Kronecker product.}
\label{tab: titan_overall}
\end{table*}

\begin{table*}[t]
\centering
\small
\begin{adjustbox}{width=0.8\linewidth}
\begin{tabular}{lccccc}
\hline
\textbf{Multimodal exp.} & \textbf{BRCA} & \textbf{GBMLGG} & \textbf{NSCLC} & \textbf{RCC} & \textbf{Overall} \\
\hline
\multicolumn{6}{c}{Cancer Subtype Prediction} \\
\hline
\textbf{ModalTune} &  $0.899 \pm 0.026$ &  $\textbf{1.000} \pm 0.000$ &  $0.956 \pm 0.010$ &  $\textbf{0.959} \pm 0.003$ &  $\textbf{0.954}$\\
\textbf{ModalTune w/ Clinical} & $\textbf{0.904} \pm 0.020$ & $0.998 \pm 0.003$ & $\textbf{0.959} \pm 0.001$ & $0.938 \pm 0.010$ & $0.950$ \\
\hline
\multicolumn{6}{c}{Survival Prediction} \\
\hline
\textbf{ModalTune} & $0.772 \pm 0.008$ & $0.879 \pm 0.004$ & $0.608 \pm 0.023$ & $0.743 \pm 0.004$ & $0.750$\\
\textbf{ModalTune w/ Clinical} & $\textbf{0.777} \pm 0.012$ & $\textbf{0.885} \pm 0.013$ & $\textbf{0.609} \pm 0.016$ & $\textbf{0.748} \pm 0.019$ & $\textbf{0.755}$ \\
\hline
\end{tabular}
\end{adjustbox}
\caption{Experiments with incorporating clinical data alongside transcriptomics in ModalTune. Best model in \textbf{bold}.}
\label{tab: supp_table_multimodal}
\end{table*}

\begin{table}[]
\centering
\small
\begin{adjustbox}{width=0.82\linewidth}
\begin{tabular}{lccccc}
\hline
 & \textbf{COADREAD} & \textbf{BLCA} \\
\hline
\multicolumn{3}{c}{\textbf{Cancer Subtype Prediction}} \\
\hline
\textbf{TITAN LP} & $0.556$ & $0.675$ \\
\textbf{TITAN Sup. (cat)} & $ 0.585\pm0.026 $ & $ 0.694 \pm 0.013 $ \\
\hline
\textbf{TITAN Cls. (cat)} & $0.511 \pm 0.018$ & $0.526 \pm 0.044$ \\
\textbf{TITAN Surv. (cat)} & $\underline{0.522} \pm 0.020$ & $\underline{0.597} \pm 0.018$ \\
\textbf{ModalTune TITAN} & $\textbf{0.583} \pm 0.089$ & $\textbf{0.691} \pm 0.016$ \\
\hline
\multicolumn{3}{c}{\textbf{Survival Prediction}} \\
\hline
\textbf{TITAN LP} & $0.562$ & $0.615$ \\
\textbf{TITAN Sup. (cat)} & $ 0.593\pm0.036 $ & $ 0.679\pm0.015 $ \\
\hline
\textbf{TITAN Cls. (cat)} & $0.483 \pm 0.038$ & $\textbf{0.617} \pm 0.017$ \\
\textbf{TITAN Surv. (cat)} & $\underline{0.549} \pm 0.051$ & $0.609 \pm 0.032$ \\
\textbf{ModalTune TITAN} & $\textbf{0.581} \pm 0.062$ & $\underline{0.611} \pm 0.063$ \\
\hline
\end{tabular}
\end{adjustbox}
\caption{Generalization study on OOD datasets using different TITAN-based models, compared with TITAN Sup. (cat) trained directly on the OOD data. Best OOD model in \textbf{bold}, second best is \underline{underlined}.}
\label{tab: titan_generalizability}
\end{table}

\section{Ablation Studies}
\label{sec:supp_ablations}
In this section, we investigate multiple key design choices in our study: impact of different Modal Adapters, the effect of Modal Adapters, the effect of text embeddings, the impact of training solely on general prompts, the impact of different text encoders and Projectors, illustrated in \cref{tab: ablations}.

\subsection{Modal Adapters}

\textbf{Different Modal Adapters (LoRA):} We compared ModalTune by extending LoRA \cite{hu2022lora}, instead of using the ViT Adapter–based Modal Adapter, to handle transcriptomics and its interactions with the slide encoder. Overall, we observed that LoRA slightly underperformed ModalTune in both cancer subtype classification (1.5\% drop) and survival prediction (0.8\% drop), thereby motivating our choice of the Modal Adapter architecture.

\textbf{Effect of Modal Adapters:} To assess more deeply if the Modal Adapter architecture provides benefits in uni-modal fine-tuning, we evaluate the performance of ModalTune by replacing transcriptomics tokens from the genomic encoder with the same number and dimension of randomly initialized \textit{trainable} embedding vectors (`Single Modal'). We find that overall, the model outperforms all other image-only models (\cref{tab: diagnosis_and_survival}) in subtype classification and has competitive performance in survival prediction. This effect is most pronounced when compared against the Gigapath fully fine-tuned model, where the model demonstrates superior performance across cancer subtype classifications and survival prediction (0.9\%; 3.4\%). Fine-tuning with the Modal Adapter setup requires updating fewer parameters than fully tuning Gigapath.  Thus, this experiment demonstrates both the efficiency and effectiveness of the proposed architecture.
\subsection{Multi-task using Texts}
\label{sec:supp_mtl}
To examine the effect of multi-task learning, we train the Modal Adapter in a single-task manner, without embedding the tasks using text (`No Text Embedding'). We found cancer subtype prediction (0.6\% drop) was less affected than survival prediction (2.7\% drop), indicating the latter utilized \textit{more} information from other tasks than the former. These findings indicate the utility of using text embeddings for multi-task learning and suggest that inter-task information is beneficial for downstream performance.

We also tested the utility of text embeddings for multi-task learning on other architectures such as ABMIL (cat) (`ABMIL (cat) w/ text emb.'), and found that compared to ABMIL (cat) trained on single tasks, there was an overall drop in performance for cancer subtype classification (1.6\% drop), while performance for survival prediction remained similar. However, better architectures like ModalTune, incorporating interaction terms, were able to achieve substantially improved performance when trained with text embeddings.

\subsection{Task-Prompts}
Here we investigate the role of using a multi-task prompt formulation versus simply pooling all tasks together into a general prompt, and performing single-task training. We do so by comparing our baseline model trained using both general and task-specific text embeddings ($T=3$, `ModalTune') versus a model trained solely on a general prompt ($T=1$, `Single Task Prompt'). Our results indicate that training with a single task prompt worsens overall model performance (2.0\% drop in subtype prediction, 1.5\% drop in survival prediction), potentially due to the regularization effects introduced by additional constraints that maximize the KL divergence between individual task-specific text vectors.
\subsection{Text encoders}
To evaluate the performance of ModalTune when using a different text embedding LLM, we tested Llama-3-8B-UltraMedical \cite{zhang2024ultramedical}. We observed a major drop in performance compared to ModalTune (3.4\% in subtype prediction, 1.5\% in survival prediction). We hypothesize several reasons for this decline. First, Llama-3-8B-UltraMedical is a general-purpose model trained on large-scale medical text datasets, whereas CONCH was trained in a contrastive manner using histopathology-related text datasets against image patches. This specialized training likely made CONCH a better fit for our use case, leading to superior performance. Similar findings were also reported in \cite{mahmoodl9:online}, where specialized models outperformed the generic model in the molecular status prediction task. Additionally, Llama-3-8B-UltraMedical is a generative model, requiring mean-pooling after encoding to obtain a single $4096$-dimensional text representation. In contrast, CONCH directly outputs a more compact text representation ($512$-dimensions), which may reduce the chance of overfit and hence improve ModalTune training.
\subsection{Projectors}
We found ModalTune to perform best when using a frozen and randomly-initialized Projector (`ModalTune'), which we explore here. Removing the Projector (`No Projector') simply requires adjusting the final output dimension $D_{final}$ to 512, matching the dimensionality of text embeddings, $D_{text}$. This adjustment resulted in a drop in performance (0.7\% in cancer subtype, 2.7\% in survival prediction). We expect this occurred because the noise introduced by the random Projector has a regularizing effect on training, reducing model overfit on specific cancer sites. This has also been explored by Arani et. al. \cite{arani2021noise}, where the introduction of noise in the knowledge distillation framework had positive effects. We additionally explore \textit{training} the randomly-initialized Projector (`Trainable Projector'), which results in severe degradations in performance on both tasks. We believe this is due to model collapse, where the KL divergence loss function could be easily minimized by having the projector and the Modal Adapter output trivial solutions. The impact on survival prediction is less pronounced, which we attribute to the C-index metric being dependent only on \textit{relative} ordering of risk scores. To avoid model collapse while tuning the Projector, we attach it to the end of the Modal Adapter instead of the text embeddings (`Model-side Projector'). We found best results when using a trained linear projection, though it still results in slightly inferior performance. 

While unorthodox, these findings do align with prior studies highlighting the utility of randomly-initialized and fixed projectors in extracting non-trivial features in various scenarios. Of particular relevance to ModalTune, random projectors are effective feature extractors, reducing dimensionality and producing powerful representations \cite{Zhao2023DatasetMatching, Amid2022LearningFeatures, saxe2011random}. Additionally, random projectors largely preserve inter-sample distances, as discussed in \cite{giryes2016deep, Zhao2023DatasetMatching}, i.e., they maintain smaller distances between samples of the same class and larger distances between samples from different classes. This is evident empirically through the performance of linear regression on text embeddings (\cref{tab: textembeddingperformance}) with and without random projectors and theoretically from the Johnson-Lindenstrauss lemma \cite{johnson1984extensions}, as discussed in Boutsidis et al. \cite{boutsidis2010random}. Given that clusters in the dataset are largely preserved and we opt to perform simple linear probing on extracted features, we expect the fixed random Projector to be a viable, generalizable, and effective projection method for ModalTune.

Overall, we find that using Modal Adapters, combining tasks with a text embedding, using multiple task prompts, and employing a fixed, randomly initialized Projector are all key components of ModalTune's success in improving the fine-tuning of SLFMs.

\section{Experiments with TITAN}
\label{sec:supp_titan}

To demonstrate the ability of ModalTune to extend to other Transformer-based SLFMs, we perform experiments interfacing it with the TITAN \cite{Ding2024MultimodalPathology} SLFM. To do so, we first re-extract patch features using the method described in the original work. We then perform analogous comparisons to Gigapath-ModalTune (\cref{tab: titan_overall}). We found TITAN to be a much stronger standalone model than Gigapath, obtaining strong results with only linear probing or fine-tuning. Nonetheless, we find TITAN benefits from the bulk transcriptomics modality over uni-modal fine-tuning (1.3\% in subtype classification, 1.0\% in survival prediction), and slightly improves when tuned using the ModalTune pipeline (1.1\%, 0.4\%). 

We additionally probe the generalizability of ModalTune TITAN similarly to the generalization study in \cref{sec:generalization_study} (\cref{tab: titan_generalizability}). We found the ModalTune framework to greatly benefit OOD prediction performance on COADREAD and BLCA, with an average of 13.9\% improvement in subtype prediction and 2.9\% improvement in survival prediction than the next best OOD baseline. Furthermore, ModalTune demonstrated generalizability, performing only 0.4\% worse than a fully-supervised TITAN (cat) network in subtype prediction and 6.3\% worse in survival prediction. Thus, even though we obtained modest improvements from in-domain validation, we emphasize ModalTune maintains SLFM generalization better than conventional tuning methods.

\section{Additional Modalities}
\label{sec:supp_multimodal}
To validate ModalTune’s extensibility, we integrated clinical data ($m_2$) for TCGA by incorporating available features: patient age (all); TNM staging (BRCA, NSCLC, RCC); treatment type (BRCA only); and hormone receptor status (BRCA only). A 2-layer MLP encodes $m_2$ into $\mathbb{R}^{1 \times D}$ for concatenation with transcriptomics [\cref{sec:modal_enc}]. Overall, in our experiments shown in \cref{tab: supp_table_multimodal}, the addition of clinical data marginally improved ModalTune’s performance, primarily for survival prediction, while in the case of RCC subtype classification, it even led to a degradation, possibly because the clinical features for RCC are less relevant for the subtype classification task. This highlights ModalTune’s ability to integrate salient features across modalities, though incorporating additional modalities remains a direction for future work.

\section{Qualitative Analysis}
\label{sec:supp_qualanalysis}
\subsection{t-SNE Analysis}
\label{sec:supp_TSNEanalysis}
\begin{figure*}[t]
  \centering
   \includegraphics[width=1.0\linewidth]{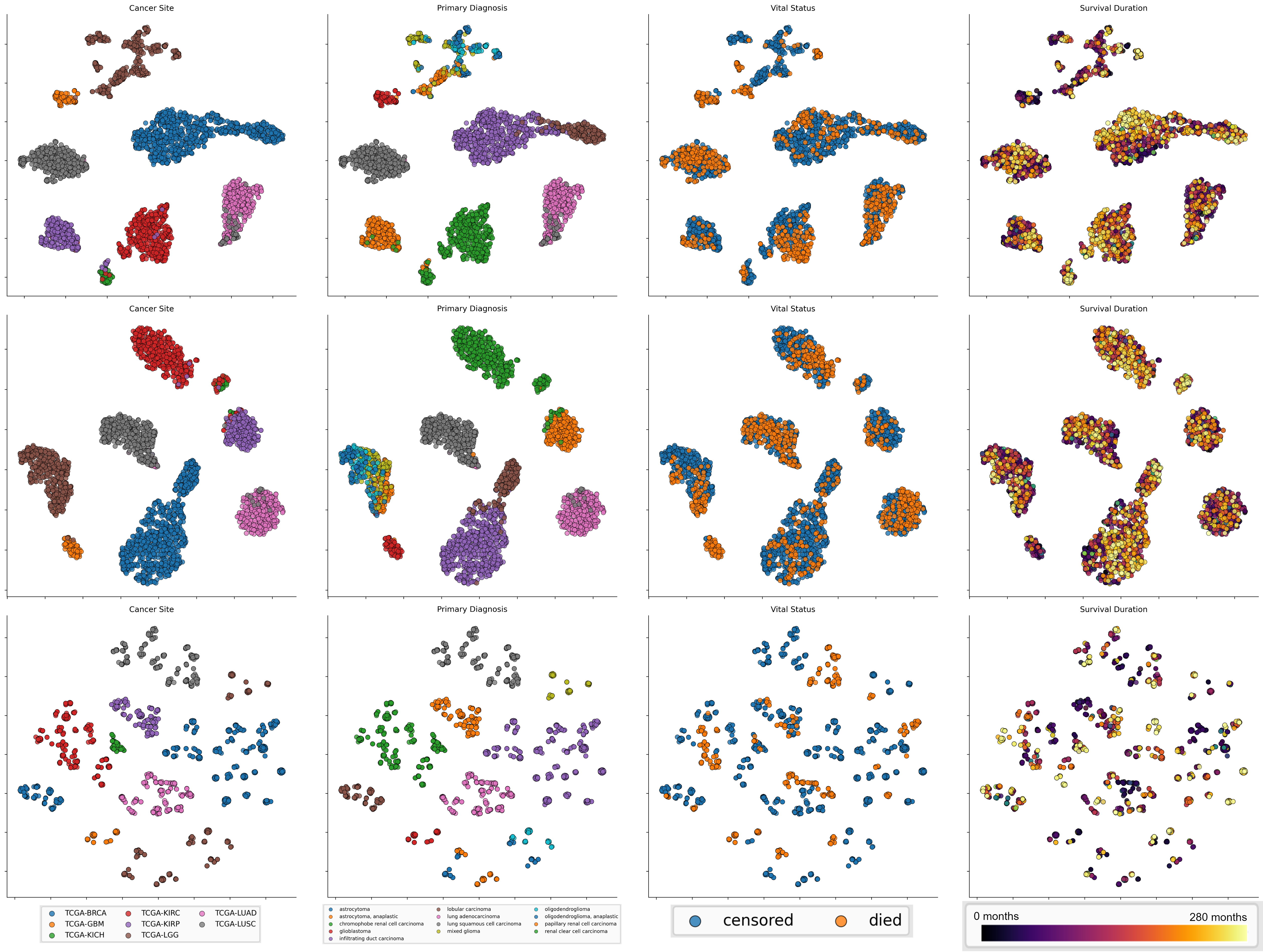}
   \caption{t-SNE plots generated for embeddings extracted using ModalTune (\textbf{first row}), ModalTune Pan-Cancer (\textbf{second row}), and text embedding vectors (\textbf{third row}). From left to right, each data point is colored by \textbf{cancer site}, \textbf{primary diagnosis}, \textbf{vital status}, and \textbf{survival duration}. }
   \label{fig:TSNEplots}
\end{figure*}

After training ModalTune and ModalTune Pan-Cancer, we extract embedding vectors from combined train, validation, and testing datasets for every cancer site. For standard ModalTune, we extract embeddings using the best model per cancer site. For ModalTune Pan-Cancer, we simply use the overall best model. t-SNE plots of the extracted embeddings, along with text embeddings, are visualized in \cref{fig:TSNEplots}. 

Notably, regardless of cancer sites being trained separately or together in a pan-cancer setup, embedding vectors distinctly cluster into individual sites. This may partially explain why we found minimal benefit in in-domain datasets from the pan-cancer experiments, as there is not much shared information between sites. We see much better separation in the former when comparing GBMLGG for standard ModalTune versus ModalTune Pan-Cancer. We expect this is due to issues with convergence mentioned in \cref{sec:quant_analysis}, where the best pan-cancer model had not yet converged on GBMLGG. In all other cases, embeddings are clearly clustered into groups based on primary diagnosis. In contrast, while text embeddings remain well separated for vital status and survival duration, separation is not as clear for embeddings from ModalTune. This is likely due to the inherently noisy nature of survival prediction. Since text prompts are directly created from clinical data and can only take discretized values, text embeddings are markedly more sparse than those generated from ModalTune. 

\subsection{Kaplan Meier Analysis}
\label{sec:supp_kaplanmeier}
\begin{figure*}[t]
  \centering
   \includegraphics[width=1.0\linewidth]{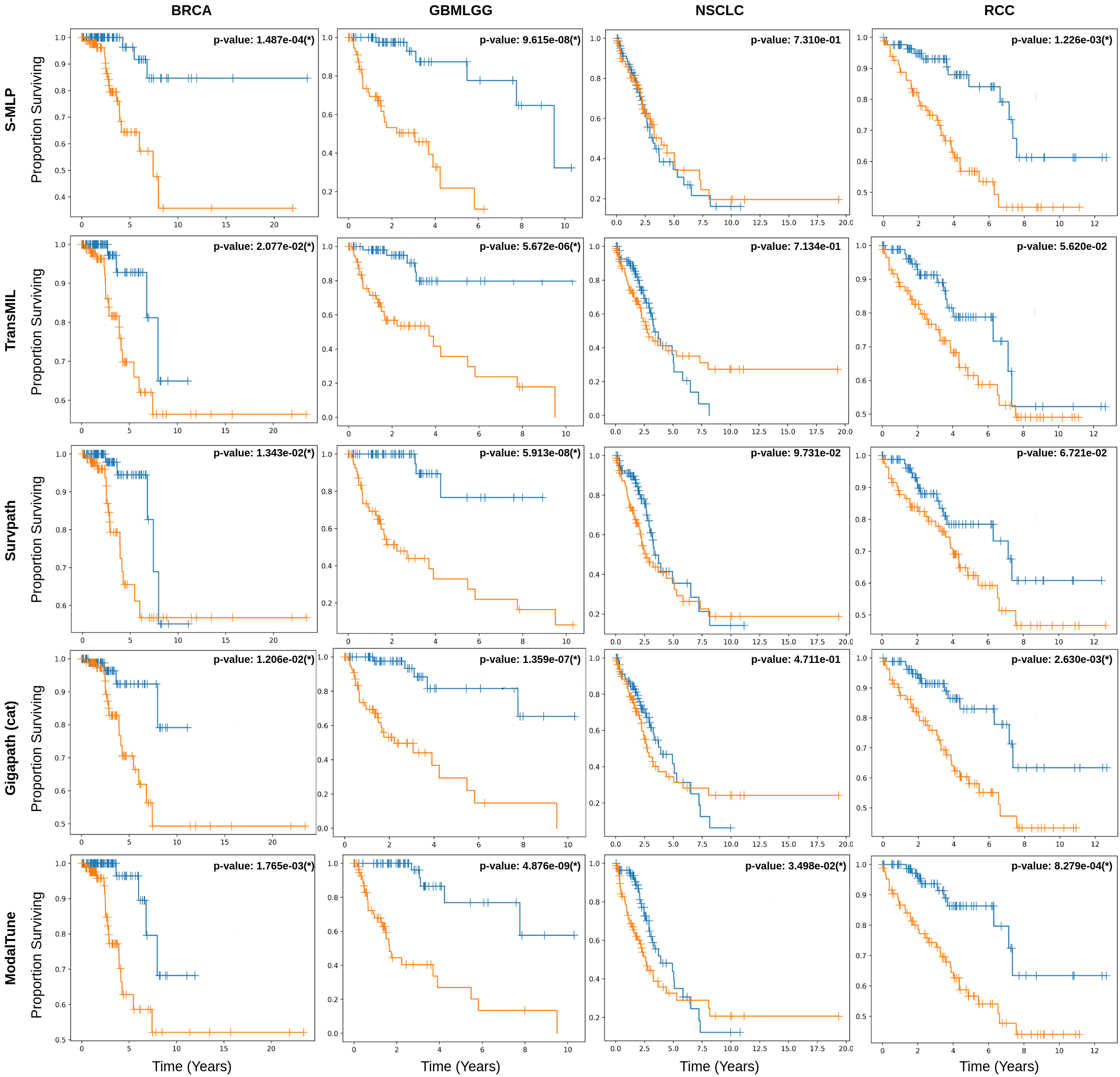}
   \caption{Kaplan-Meier curves of ModalTune and the best-performing survival model baselines across four cancer types. Patient groups are stratified based on the median of model-estimated risk scores on the test set, with orange representing the low-risk group and blue representing the high-risk group. A log-rank test with a significance threshold of $\alpha=0.05$ was used to assess differences between the two distributions}
   \label{fig:kaplan_meier}
\end{figure*}

Although a high c-index risk model is preferred, it is equally important for the model to stratify patients into two distinct groups to aid clinicians in making treatment decisions, allowing them to choose between more or less aggressive interventions based on the patient's risk group. We used Kaplan-Meier curves on the test set to visualize this stratification, comparing high-risk and low-risk groups. The two groups were then assessed using a log-rank test to measure differences between their survival distributions, with a significance threshold set at $\alpha=0.05$. In \cref{fig:kaplan_meier}, we compare ModalTune against the best-performing survival models from image-only, genomics-only, and multi-modal categories, as well as Gigapath (cat). ModalTune consistently maintains significance in patient stratification across all four cancer types. ModalTune is the only model whose stratification was significant for NSCLC. Interestingly, the Kaplan-Meier curves for both ModalTune and Gigapath (cat) show strong similarities in patient stratification and their pattern, with ModalTune achieving improved stratification through better integration of transcriptomics information.

%

\end{document}